\documentclass[acmsmall,nonacm]{acmart}
\pdfoutput=1

\acmJournal{PACMPL}
\acmVolume{1}
\acmNumber{CONF} 
\acmArticle{1}
\acmYear{2018}
\acmMonth{1}
\acmDOI{} 
\startPage{1}

\setcopyright{none}

\usepackage{amsmath,amssymb,amsthm}
\usepackage{tikz-cd}
\usepackage{array}
\usepackage{tikz}
\usepackage{pgfplots}

\usepackage{enumitem}

\usepackage{clrscode3e}
\usepackage{hyperref}
\hypersetup{
    colorlinks,
    linkcolor={blue!70!black},
    citecolor={blue!70!black},
    urlcolor={blue!70!black}
}
\usepackage{todonotes}
\usepackage{thmtools}
\usepackage{thm-restate}
\usepackage{eucal}
\usepackage{calc}
\usepackage{adjustbox}
\usepackage{wrapfig}
\usepackage{cutwin}
\usepackage{microtype}
\usepackage{multirow,bigdelim}
\usepackage{mathtools}
\usepackage{subfig}
\usepackage{mathrsfs}
\usepackage{chngcntr}
\usepackage{apptools}
\usepackage{booktabs}

\AtAppendix{\counterwithin{theorem}{section}}
\AtAppendix{\counterwithin{theorem}{section}}

\usetikzlibrary{cd,automata}

\theoremstyle{plain}
\newtheorem{theorem}{Theorem}[section]
\newtheorem{proposition}[theorem]{Proposition}
\newtheorem{lemma}[theorem]{Lemma}

\theoremstyle{definition}
\newtheorem{remark}[theorem]{Remark}

\newcommand{\cat}[1]{\mathbf{#1}}

\newcommand{\Set}{\cat{Set}}
\newcommand{\idf}{\mathsf{id}}
\newcommand{\Id}{\mathsf{Id}}

\renewcommand{\gets}{\leftarrow}

\newcommand{\Ps}{\mathcal{P}}

\newcommand{\ev}{\mathsf{ev}}
\newcommand{\img}{\mathtt{img}}
\newcommand{\eword}{\varepsilon}
\newcommand{\pref}{\mathsf{prefixes}}

\newcommand{\lstar}{\ensuremath{\texttt{L}^{\!\star}}}
\newcommand{\nlstar}{\ensuremath{\texttt{NL}^{\!\star}}}

\newcommand{\lstarT}[1][T]{\ensuremath{\lstar_{#1}}}

\newcommand{\lang}{\mathcal{L}}

\newcommand{\row}{\mathsf{row}}
\newcommand{\Trow}{\row_{\mathsf{t}}}

\newcommand{\Brow}{\mathsf{row}_{\mathsf{b}}}

\newcommand{\init}{\mathsf{init}}
\newcommand{\out}{\mathsf{out}}

\newcommand{\autom}{\mathcal{A}}
\newcommand{\hyp}{\mathcal{H}}
\newcommand{\suc}{\mathcal{S}}
\newcommand{\minaut}{\mathcal{M}}

\newcommand{\N}{\mathbb{N}}

\newcommand{\MMon}{\mathtt{Writer}}
\newcommand{\mon}{\mathbb{M}}

\newcommand{\sring}{\mathbb{S}}
\newcommand{\VMon}{V}

\newenvironment{automaton}{\begin{tikzpicture}[node distance=1.5cm,on grid,auto,initial text={},state/.append style={inner sep=0pt,outer sep=0pt,minimum size=3.5ex}]}{\end{tikzpicture}}

\newcolumntype{C}[1]{>{\centering\arraybackslash$}m{#1}<{$}}
\newcolumntype{L}[1]{>{\raggedleft\arraybackslash$}m{#1}<{$}}
\newcolumntype{R}[1]{>{\raggedright\arraybackslash$}m{#1}<{$}}

\def\definitionname{Definition}
\def\examplename{Example}

\begin{document}

\title{Optimizing automata learning via monads}

\author{Gerco van Heerdt}
\email{gerco.heerdt@ucl.ac.uk}
\author{Matteo Sammartino}
\email{m.sammartino@ucl.ac.uk}
\author{Alexandra Silva}
\email{alexandra.silva@ucl.ac.uk}
\affiliation{
  \institution{University College London}
}

\begin{abstract}
	Automata learning has been successfully applied in the verification of hardware and software. The size of the automaton model learned is a bottleneck for scalability, and hence optimizations that enable learning of compact representations are important. This paper exploits monads, both as a mathematical structure and a programming construct, to design, prove correct, and implement a wide class of such optimizations. The former perspective on monads allows us to develop a new algorithm and accompanying correctness proofs, building upon a general framework for automata learning based on category theory. The new algorithm is parametric on a monad, which provides a rich algebraic structure to capture non-determinism and other side-effects.  We show that our approach allows us to uniformly capture existing algorithms, develop new ones, and add optimizations. The latter perspective allows us to effortlessly translate the theory into practice: we provide a Haskell library implementing our general framework, and we show experimental results for two specific instances: non-deterministic and weighted automata.
	
	
\end{abstract}


\maketitle

\section{Introduction}



The increasing complexity of software and hardware systems calls for new scalable methods to design, verify, and continuously improve systems. Black-box inference methods aim at building models of running systems by observing their response to certain queries. This reverse engineering process is very amenable for automation and allows for fine-tuning the precision of the model depending on the properties of interest, which is important for scalability. 

One of the most successful instances of black-box inference is automata learning, which has been used in various verification tasks, ranging from finding bugs in implementations of network protocols~\citep{ruiter2015} to rejuvenating legacy software~\citep{schuts2016}. \citet{cacm} has recently written a comprehensive overview of the widespread use of automata learning in verification.


A limitation in automata learning is that the models of real systems can become too large to be handled by tools. This demands compositional methods and techniques that enable compact representation of behaviors.

In this paper, we show how monads can be used to add optimizations to learning algorithms in order to obtain compact representations. We will use as playground for our approach the well known \lstar\ algorithm~\citep{angluin1987}, which learns a minimal deterministic finite automaton (DFA) accepting a regular language by interacting with a \emph{teacher}, i.e., an oracle that can reply to specific queries about the target language. Monads allow us to take an abstract approach, in which category theory is used to devise an optimized learning algorithm and a generic correctness proof for a broad class of compact models. 
Monads also allow us to straightforwardly implement the algorithm in Haskell via the corresponding programming constructs.

The inspiration for this work 
is quite concrete: it is a well-known fact that non-deterministic finite automata (NFAs) can be much smaller than deterministic ones for a regular language. The subtle point is that given a regular language, there is a canonical deterministic automaton accepting it---the minimal one---but there might be many ``minimal'' non-deterministic automata accepting the same language. This raises a challenge for learning algorithms: which non-deterministic automaton should the algorithm learn? To overcome this, \citet{bollig2009} developed a version of Angluin's \lstar\ algorithm, which they called \nlstar, in which they use a particular class of NFAs, namely \emph{Residual Finite State Automata} (RFSAs), which do admit minimal canonical representatives. Though \nlstar\ indeed is a first step in incorporating a more compact representation of regular languages, there are several questions that remain to be addressed.
We tackle them in this paper.

DFAs and NFAs are formally connected by the subset construction.
Underpinning this construction is the rich algebraic structure of languages and of the state space of the DFA obtained by determinizing an NFA.
The state space of a determinized DFA---consisting of subsets of the state space of the original NFA---has a join-semilattice structure. Moreover, this structure is preserved in language acceptance: if there are subsets $U$ and $V$, then the language of $U \cup V$ is the union of the languages of the first two.
Formally, the function that assigns to each state its language is a join-semilattice map, since languages themselves are just sets of words and have a lattice structure.
And languages are even richer: they have the structure of complete atomic Boolean algebras.
This leads to several questions: Can we exploit this structure and have even more compact representations? What if we slightly change the setting and look at weighted languages over a semiring, which have the structure of a semimodule (or vector space, if the semiring is a field)? 

The latter question is strongly motivated by the widespread use of weighted languages and corresponding \emph{weighted finite automata} (WFAs) in verification, from the formal verification of quantitative properties~\citep{ChatterjeeDH08,DrosteG05,Kuperberg14}, to probabilistic model-checking~\citep{BaierGC09}, to the verification of on-line algorithms~\citep{AminofKL10}. 

Our key insight is that the algebraic structures mentioned above are in fact algebras for a monad $T$. In the case of join-semilattices this is the powerset monad, and in the case of vector spaces it is the free vector space monad. These monads can be used to define a notion of $T$-automaton, with states having the structure of an algebra for the monad $T$,
which generalizes non-determinism as a side-effect. From $T$-automata we can derive a compact, equivalent version by taking as states a set of {\em generators} and transferring the algebraic structure of the original state space to the transition structure of the automaton.

This general perspective enables us to generalize $\lstar{}$ to a new algorithm $\lstarT{}$, which learns compact automata featuring non-determinism and other side-effects captured by a monad. Moreover, $\lstarT{}$ incorporates further optimizations arising from the monadic representation, which lead to more scalable algorithms.

Besides the theoretical aspects, we devote large part of this paper to implementation and experimental evaluation.
Monads are key for us to faithfully translate theory into practice. We provide a library that implements all aspects of our general framework, making use of Haskell monads.\footnote{The code is provided as supplementary material.} For any monad, the library allows the programmer to obtain a basic, correct-by-construction instance of the algorithm and of its optimized versions for free. This enables the programmer to experiment with different optimizations with minimal effort. Our library also allows the programmer to re-define some basic operations, if a more efficient version is available, in order to make the algorithm more amenable to real-world usage. For instance, generators can be computed efficiently in the vector space case via Gaussian elimination. 



One of the main challenges in applying Angluin-style algorithms to real-world systems is implementing the teacher. In fact, it is often the case that exact answers to certain queries are not available. In these cases the teacher often resorts to \emph{random testing}~\citep[e.g.,][]{aarts2013,cho2010,chalupar2014}, with an unavoidable trade-off in terms of model accuracy (see~\citep{cacm} for a detailed discussion on this issue). Our library provides support for both exact and approximate teachers, along with a basic implementation that works for any monad. Interestingly, the exact teacher relies on \emph{bisimulation up to context}~\citep{sangiorgi1998,rot2013}, which exploits the monad structure to efficiently determine bisimulation.

\section{Overview and contributions}\label{sec:overview}

In this section, we give an overview of our approach and highlight our main contribution. We start by explaining the original \lstar{} algorithm. We then  discuss the challenges in adapting the algorithm to learn automata with side-effects, illustrating them through a concrete example---NFAs. 



\subsection{\lstar{} algorithm}

The \lstar{} algorithm learns the minimal DFA accepting a language $\lang \subseteq A^*$ over a finite alphabet $A$. The algorithm assumes the existence of a \emph{minimally adequate teacher}, which is an oracle that can answer two types of queries:
\begin{itemize}
	\item
		\textbf{Membership queries}: given a word $w \in A^*$, does $w$ belong to $\lang$?
	\item
		\textbf{Equivalence queries}: given a \emph{hypothesis} DFA $\hyp$, does $\hyp$ accept $\lang$?
		If not, the teacher will return a \emph{counterexample}, i.e., a word incorrectly accepted or rejected by $\hyp$.
\end{itemize}
The algorithm incrementally builds an \emph{observation table}. 
The table is made of two parts: a top part, with rows ranging over a finite set $S \subseteq A^*$; and a bottom part, with rows ranging over $S \cdot A$ (i.e., words of the form $s a$, with $s \in S$ and $a \in A$). Columns range over a finite $E \subseteq A^*$. For each $u \in S \cup S \cdot A$ and $v \in E$, the corresponding cell in the table contains $1$ if and only if $uv \in \lang$. Intuitively, each row $u$ contains enough information to fully identify the Myhill-Nerode equivalence class of $u$ with respect to an approximation of the target language---rows with the same content are considered members of the same equivalence class. Cells are filled in by the algorithm via membership queries.




%


 As an example, and to set notation, consider the table below over $A = \{a,b\}$.
It shows that $\lang$ contains the word $aa$ and does not contain the words $\eword$ (the empty word), $a$, $b$, $ba$, $aaa$, and $baa$.
\begin{minipage}[c]{.3\linewidth}
	\centering
	\begin{tabular}{L{3em} R{2ex} | C{2ex}C{2ex}C{3ex}@{}m{0pt}@{}}
		\multicolumn{5}{c}{$\hspace{13ex}\overbracket[.8pt][2pt]{\rule{12ex}{0pt}}^{\displaystyle E}$} \\
		& & \eword &  a & aa
		\\
		\cline{2-5}
		\ldelim[{1}{5mm}[$S$] & \eword & 0 & 0 & 1&
		\\[1ex]
		\cline{2-5}
		\ldelim[{2}{10mm}[$S \cdot A$]
		& a & 0 & 1 & 0 &
		\\
		& b & 0 & 0 & 0 &
	\end{tabular}
\end{minipage}
\begin{minipage}[c]{.7\linewidth}
	\begin{alignat*}{4}
		\Trow &\colon S \to 2^E &\Trow(u)(v) &= 1
		&&\iff &uv \in \lang
		\\
		\Brow &\colon S \to (2^E)^A
		&
		\quad \Brow(u)(a)(v) &= 1 &&\iff &uav \in \lang
	\end{alignat*}
\end{minipage}
%
We use the functions $\Trow$ and $\Brow$ to describe the top and bottom parts of the table, respectively. Notice that $S$ and $S \cdot A$ may intersect. For the sake of conciseness, when tables are depicted, elements in the intersection are only shown in the top part.
%
%
%

A key idea of the algorithm is to construct a hypothesis DFA from the different rows in the table. The construction is the same as that of the minimal DFA from the Myhill-Nerode equivalence, and exploits the correspondence between table rows and Myhill-Nerode equivalence classes. The state space of the hypothesis DFA is given by the set $H = \{\Trow(s) \mid s \in S\}$.
Note that there may be multiple rows with the same content, but they result in a single state, as they all belong to the same Myhill-Nerode equivalence class.
The initial state is $\Trow(\eword)$, and we use the $\eword$ column to determine whether a state is accepting: $\Trow(s)$ is accepting whenever $\Trow(s)(\eword) = 1$.
The transition function is defined as $\Trow(s) \xrightarrow{a} \Brow(s)(a)$.
(Notice that the continuation is drawn from the bottom part of the table).
For the hypothesis automaton to be well-defined, $\eword$ must be in $S$ and $E$, and the table must satisfy two properties:
\begin{itemize}
	\item
		\textbf{Closedness} states that each transition actually leads to a state of the hypothesis.
		That is, the table is closed if for all $t \in S$ and $a \in A$ there is $s \in S$ such that $\Trow(s) = \Brow(t)(a)$.
	\item
		\textbf{Consistency} states that there is no ambiguity in determining the transitions.
		That is, the table is consistent if for all $s_1, s_2 \in S$ such that $\Trow(s_1) = \Trow(s_2)$ we have $\Brow(s_1) = \Brow(s_2)$.
\end{itemize}
The algorithm updates the sets $S$ and $E$ to satisfy these properties, constructs a hypothesis, submits it in an equivalence query, and, when given a counterexample, refines the hypothesis.
This process continues until the hypothesis is correct.
The algorithm is shown in \figurename~\ref{fig:lstar}.


\begin{figure}[t]
	\begin{codebox}
		\li $S, E \gets \{\eword\}$
		\li \Repeat
			\li \While the table is not closed or not consistent
				\li \Do \If the table is not closed
					\li\label{line:closedness} \Then find $t \in S,a \in A$ such that $\Brow(t)(a) \neq \Trow(s)$ for all $s \in S$
					\li $S \gets S \cup \{ta\}$
				\End
				\li \If the table is not consistent
					\li\label{line:consistency} \Then find $s_1, s_2 \in S$, $a \in A$, and $e \in E$ such that
					\zi\qquad $\Trow(s_1) = \Trow(s_2)$ and $\Brow(s_1)(a)(e) \neq \Brow(s_2)(a)(e)$
					\li $E \gets E \cup \{ae\}$
				\End
			\End
			\li\label{line:hypothesis}Construct the hypothesis $\hyp$ and submit it to the teacher
			\li \If the teacher replies \emph{no}, with a counterexample $z$
			\li \Then $S \gets S \cup \pref(z)$
		\End
		\li \Until the teacher replies \emph{yes}
		\li \Return $\hyp$
	\end{codebox}
	\caption{\lstar{} algorithm.}
	\label{fig:lstar}
\end{figure}

\paragraph{Example Run.}
\captionsetup[subfigure]{justification=centering}
\begin{figure}[t]%
\vspace{-10ex}
\centering
	\subfloat[]{\label{fig:obs1}%
		\begin{tabular}[b]{r | c}
			& $\eword$ \\
			\hline
			$\eword$ & 1 \\
			\hline
			$a$ & 0
		\end{tabular}%
	}\qquad%
	\subfloat[]{\label{fig:obs2}%
		\begin{tabular}[b]{r | c}
			& $\eword$ \\
			\hline
			$\eword$ & 1 \\
			$a$ & 0 \\
			\hline
			$aa$ & 1
		\end{tabular}%
	}\qquad%
	\subfloat[]{\label{fig:hyp1}%
		\begin{automaton}
			\node[initial,state,accepting] (q0) {};
			\node[state] (q1) [right of=q0] {};
			\node (rr) [above of=q0] {};
			\node [above of=rr] {};
			\path[->]
			(q0) edge [bend left] node {$a$} (q1)
			(q1) edge [bend left] node {$a$} (q0);
		\end{automaton}
	}\qquad%
	\subfloat[]{\label{fig:obs3}%
		\begin{tabular}[b]{r | c}
			& $\eword$ \\
			\hline
			$\eword$ & 1 \\
			$a$ & 0 \\
			$aa$ & 1 \\
			$aaa$ & 1 \\
			\hline
			$aaaa$ & 1 \\
		\end{tabular}%
	}\qquad%
	\subfloat[]{\label{fig:obs4}%
		\begin{tabular}[b]{r | cc}
			& $\eword$ & $a$ \\
			\hline
			$\eword$ & 1 & 0 \\
			$a$ & 0 & 1 \\
			$aa$ & 1 & 1 \\
			$aaa$ & 1 & 1 \\
			\hline
			$aaaa$ & 1 & 1 \\
		\end{tabular}%
	}%
	\caption{Example run of \lstar{} on $\lang = \{w \in \{a\}^* \mid |w| \ne 1\}$.}%
\end{figure}
We now run the algorithm with the target language $\lang = \{w \in \{a\}^* \mid |w| \ne 1\}$. The minimal DFA accepting $\lang$ is 
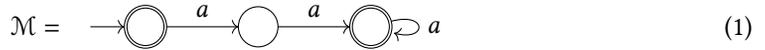
\begin{equation}
	\minaut = 
	\begin{gathered}
	\begin{automaton}
		\node[initial,state,accepting] (q0) {};
		\node[state] (q1) [right of=q0] {};
		\node[state,accepting] (q2) [right of=q1] {};
		\path[->]
		(q0) edge node {$a$} (q1)
		(q1) edge node {$a$} (q2)
		(q2) edge [loop right] node {$a$} ();
	\end{automaton}
	\end{gathered}
	\label{aut:ml}	
\end{equation}
Initially, $S = E = \{\eword\}$.
We build the observation table given in \figurename~\ref{fig:obs1}.
This table is not closed, because the row with label $a$, having 0 in the only column, does not appear in the top part of the table: the only row $\eword$ has 1.
To fix this, we add the word $a$ to the set $S$.
Now the table (\figurename~\ref{fig:obs2}) is closed and consistent, so we construct the hypothesis that is shown in \figurename~\ref{fig:hyp1} and pose an equivalence query.
The teacher replies \emph{no} and informs us that the word $aaa$ should have been accepted.
\lstar{} handles a counterexample by adding all its prefixes to the set $S$.
We only have to add $aa$ and $aaa$ in this case.
The next table (\figurename~\ref{fig:obs3}) is closed, but not consistent: the rows $\eword$ and $aa$ both have value $1$, but their extensions $a$ and $aaa$ differ.
To fix this, we prepend the continuation $a$ to the column $\eword$ on which they differ and add $a \cdot \eword = a$ to $E$.
This distinguishes $\Trow(\eword)$ from $\Trow(aa)$, as seen in the next table in \figurename~\ref{fig:obs4}.
The table is now closed and consistent, and the new hypothesis automaton is precisely the correct one $\minaut$.


As mentioned, the hypothesis construction approximates the theoretical construction of the minimal DFA, which is unique up to isomorphism.
That is, for $S = E = A^*$ the relation that identifies words of $S$ having the same value in $\Trow$ is precisely the Myhill-Nerode's right congruence.

\subsection{Learning non-deterministic automata}
\label{sec:overview-nfas}

As it is well known, NFAs can be smaller than the minimal DFA for a given language.
For example, the language $\lang$ above is accepted by the NFA
\begin{equation}
	\mathcal{N} = 
	\begin{gathered}
	\begin{automaton}
		\node[initial,state,accepting] (q0) {};
		\node[state] (q1) [right of=q0] {};
		\path[->]
		(q0) edge [bend left] node {$a$} (q1)
		(q1) edge [bend left] node {$a$} (q0)
		(q1) edge [loop right] node {$a$} ();
	\end{automaton}		
	\end{gathered}
	\label{aut:n}	
\end{equation}
%
which is smaller than the minimal DFA $\minaut$. Though in this example, which we chose for simplicity, the state reduction is not massive, it is known that in general NFAs can be exponentially smaller than the minimal DFA~\citep{kozen-book}. This reduction of the state space is enabled by a side-effect---non-determinism, in this case. 

Learning NFAs can lead to a substantial gain in space complexity, but it is challenging. The main difficulty is that
NFAs do not have a canonical minimal representative: there may be several non-isomorphic state-minimal NFAs accepting the same language, which poses problems for the development of the learning algorithm.  To overcome this, \citet{bollig2009} proposed to use a particular class of NFAs, namely RFSAs, which do admit minimal canonical representatives. However, their ad-hoc solution for NFAs does not extend to other automata, such as weighted or alternating. In this paper we present a solution that works for any side-effect, specified as a monad.

The crucial observation underlying our approach is that the language semantics of an NFA is defined in terms of its determinization, i.e., the DFA obtained by taking sets of states of the NFA as state space. In other words, this DFA is defined over an algebraic structure induced by the powerset, namely a \emph{join semilattice} (JSL) whose join operation is set union. This automaton model does admit minimal representatives, which leads to the key idea for our algorithm: learning NFAs as automata over JSLs.
%
%
%
In order to do so, we use an extended table where rows have a JSL structure, defined as follows. The join of two rows is given by an element-wise or, and the bottom element is the row containing only zeroes. More precisely, the new table consists of the two functions
%
\[
	\Trow^\sharp \colon \Ps(S) \to 2^E
	\qquad	
	\Brow^\sharp \colon \Ps(S) \to (2^E)^A	
\]
given by $\Trow^\sharp(U) = \bigvee \{\Trow(s) \mid s \in U\}$ and $\Brow^\sharp(U)(a) = \bigvee \{\Brow(s)(a) \mid s \in U\}$.
Formally, these functions are JSL homomorphisms, and they 
 induce the following general definitions:
\begin{itemize}[leftmargin=*]
	\item
		The table is {\em closed} if for all $U \subseteq S, a \in A$ there is $U' \subseteq S$ such that $\Trow^\sharp(U') = \Brow^\sharp(U)(a)$.
	\item
		The table is {\em consistent} if for all $U_1, U_2 \subseteq S$ s.t. $\Trow^\sharp(U_1) = \Trow^\sharp(U_2)$ we have $\Brow^\sharp(U_1) = \Brow^\sharp(U_2)$.
\end{itemize}
We remark that our algorithm does not actually store the whole extended table, which can be quite large. It only needs to store the original table over $S$ because all other rows in $\Ps(S)$ are freely generated and can be computed as needed, with no additional membership queries. The only lines in \figurename~\ref{fig:lstar} that need to be adjusted are lines \ref{line:closedness} and \ref{line:consistency}, where closedness and consistency are replaced with the new notions given above. Moreover, $\hyp$ is now built from the extended table.

\paragraph{Optimizations.}

In this paper we also present two optimizations to our algorithm. For the first one, note that the state space of the hypothesis constructed by the algorithm can be very large since it encodes the entire algebraic structure. 
We show that we can extract a \emph{minimal set of generators} from the table and compute a \emph{succinct hypothesis} in the form of an automaton with side-effects, without any algebraic structure. For JSLs, this consists in only taking rows that are not the join of other rows, i.e., the \emph{join-irreducibles}. By applying this optimization to this specific case, we essentially recover the learning algorithm of \citet{bollig2009}. The second optimization is a generalization of the optimized counterexample handling method of~\citet{rivest1993}, originally intended for \lstar{} and DFAs. It consists in processing counterexamples by adding a single \emph{suffix} of the counterexample to $E$, instead of adding all prefixes of the counterexample to $S$. This can avoid the algorithm posing a large number of membership queries.
%
%
%
%
%
%


\begin{figure}[t]%
\centering	
	\subfloat[]{\label{fig:obs1r}%
		\begin{tabular}[b]{r | c}
			& $\eword$ \\
			\hline
			$\eword$ & 1 \\
			\hline
			$a$ & 0
		\end{tabular}%
	}
	\hspace{8ex}
\subfloat[]{\label{fig:hyp2full}%
	\centering
		\begin{automaton}
			\node[initial,state,accepting] (q0) {};
			\node[state] (q1) [below of=q0] {};
		\path[->]
		(q0) edge node {$a$} (q1)
		(q1) edge [loop right] node {$a$} ();
		\end{automaton}%
	}
	\hspace{8ex}
\subfloat[]{\label{fig:hyp2}%
	\centering
		\begin{automaton}
			\node[initial,state,accepting] (q0) {};
		\end{automaton}%
	}
	\hspace{8ex}
	\subfloat[]{\label{fig:obs5}%
	\centering
		\begin{tabular}[b]{r | c}
			& $\eword$ \\
			\hline
			$\eword$ & 1 \\
			$a$ & 0 \\
			$aa$ & 1 \\
			\hline
			$aaa$ & 1
		\end{tabular}%
	}
	\hspace{8ex}
	\subfloat[]{\label{fig:obs6}%
	\centering
		\begin{tabular}[b]{r | cc}
			& $\eword$ & $a$ \\
			\hline
			$\eword$ & 1 & 0 \\
			$a$ & 0 & 1 \\
			$aa$ & 1 & 1 \\
			\hline
			$aaa$ & 1 & 1
		\end{tabular}%
	}%
	\caption{Example run of the \lstar{} adaptation for NFAs on $\lang = \{w \in \{a\}^* \mid |w| \ne 1\}$.}%
\end{figure}
\paragraph{Example Revisited.}
We now run the new algorithm on the language $\lang = \{w \in \{a\}^* \mid |w| \ne 1\}$ considered earlier.
Starting from $S = E = \{\eword\}$, the observation table (\figurename~\ref{fig:obs1r}) is immediately closed and consistent.
(It is closed because we have $\Trow^\sharp(\{a\}) = \Trow^\sharp(\emptyset)$.)
This gives the JSL hypothesis shown in \figurename~\ref{fig:hyp2full}, which leads to an NFA hypothesis having a single state that is initial, accepting, and has no transitions (\figurename~\ref{fig:hyp2}).
The hypothesis is obviously incorrect, and the teacher may supply us with counterexample $aa$.
Adding prefixes $a$ and $aa$ to $S$ leads to the table in \figurename~\ref{fig:obs5}.
The table is again closed, but not consistent: $\Trow^\sharp(\{a\}) = \Trow^\sharp(\emptyset)$, but $\Brow^\sharp(\{a\})(a) = \Trow^\sharp(\{aa\}) \ne \Trow^\sharp(\emptyset) = \Brow^\sharp(\emptyset)(a)$.
Thus, we add $a$ to $E$.
The resulting table (\figurename~\ref{fig:obs6}) is closed and consistent.
We note that row $aa$ is the union of other rows: $\Trow^\sharp(\{aa\}) = \Trow^\sharp(\{\eword, a\})$ (i.e., it is not a join-irreducible), and therefore can be ignored when building the succinct hypothesis. This hypothesis has two states, $\eword$ and $a$, and indeed it is the correct one $\mathcal{N}$.

\subsection{Contributions and road map of the paper}

After some preliminary notions in Section~\ref{sec:prelim}, our main contributions are presented as follows:
\begin{itemize}[leftmargin=*]
	\item In Section~\ref{sec:algorithm}, we develop a general algorithm $\lstarT{}$, which generalizes the NFA one presented in Section~\ref{sec:overview-nfas} to an arbitrary \emph{monad} $T$ capturing side-effects, and we provide a general correctness proof for our algorithm.
	\item\label{item:rep} In Section~\ref{sec:hypotheses}, we describe the first optimization and prove its correctness.
	\item\label{item:opt}
	In Section~\ref{sec:counterexamples} we describe the second optimization. We also show how it can be combined with the one of Section~\ref{sec:hypotheses}, and how it can lead to a further small optimization, where the consistency check on the table is dropped.
	\item In Section~\ref{sec:examples} we show how \lstarT{} can be applied to several automata models, highlighting further case-specific optimizations when available.
	\item In Section~\ref{sec:implementation} we describe our library and explain in detail how it can be instantiated to NFAs and WFAs. The implementation of monads for these two cases is non-trivial, due to specific Haskell requirements. We also give efficient versions of both instances. To the best of our knowledge, we are the first ones to implement an Angluin-style learning algorithm for WFAs, and to provide optimizations for it.
	\item Finally, in Section~\ref{sec:experiments} we describe experimental results for the non-deterministic and weighted cases, comparing all the optimizations enabled by our library. In particular, for NFAs we show that the Rivest and Schapire optimization, not available to \citet{bollig2009}, leads to an improvement in the number of membership queries, as happens in the DFA case.
\end{itemize}

\section{Preliminaries}\label{sec:prelim}

In this section we define a notion of $T$-automaton, a generalization of non-deterministic finite automata parametric in a monad $T$. We assume familiarity with basic notions of category theory: functors (in the category $\Set$ of sets and functions) and natural transformations.

Side-effects and different notions of non-determinism can be conveniently captured as a \emph{monad}.   A monad $T = (T,\eta,\mu)$ is a triple consisting of an
endofunctor $T$ on $\Set$ and two natural transformations: a
\emph{unit} $\eta\colon \mathtt{Id} \Rightarrow T$ 
and a \emph{multiplication} $\mu\colon
T^2 \Rightarrow T$,
which satisfy the compatibility laws
$
\mu \circ \eta_T = \idf_T = \mu \circ T\eta
$
and
$
\mu \circ \mu_T = \mu \circ T\mu$.
\begin{example}[Monads]
An example of a monad is the triple $(\Ps, \{-\}, \bigcup)$, where $\Ps$ denotes the powerset functor associating a collection of subsets to a set, $\{-\}$ is the singleton operation, and $\bigcup$ is just union of sets. Another example is the triple $(\VMon(-), e, m)$, where $\VMon(X)$ is the free semimodule (over a semiring $\sring$) over $X$, namely
$
\{
\varphi \mid \varphi \colon X \to \sring \text{ having finite support}
\}$.
	The support of a function $\varphi \colon X \to \sring$ is the set of $x \in X$ such that $\varphi (x) \neq 0$. Then $e\colon X \to \VMon(X)$ is the characteristic function for each $x\in X$, and $m\colon \VMon(\VMon(X)) \to \VMon(X)$ is defined for $\varphi \in \VMon(\VMon(X))$ and $x\in X$ as $m(\varphi)(x) = \sum_{\psi \in \VMon(X)}\varphi(\psi)\times \psi(x)$.
\end{example}
Given a monad $T$, a $T$-algebra is a pair $(X,h)$ consisting of a carrier set $X$ and a function $h\colon TX \to X$  such that $h \circ \mu_X = h \circ Th$
and $h \circ \eta_X = \idf_X$. A $T$-homomorphism between two $T$-algebras $(X,h)$ and $(Y,k)$
is a function $f\colon X \to Y$ such that $f \circ h = k \circ Tf$. The abstract notion of $T$-algebra instantiates to expected notions, as illustrated in the following example.


\begin{example}[Algebras for a monad]
The $\Ps$-algebras are the (complete) join-semilattices, and their homomorphisms are join-preserving functions. If $\sring$ is a field, $\VMon$-algebras are vector spaces, and their homomorphisms are linear maps.
\end{example}
We will often refer to a $T$-algebra $(X, h)$ as $X$ if $h$ is understood or if its specific definition is irrelevant.
Given a set $X$, $(TX, \mu_X)$ is a $T$-algebra called the \emph{free $T$-algebra} on $X$. One can build algebras pointwise for some operations. For instance, if $Y$ is a set and $(X, x)$ a $T$-algebra, then we have a $T$-algebra $(X^Y, f)$, where $f \colon T(X^Y) \to X^Y$ is given by $f(W)(y) = (x \circ T(\ev_y))(W)$ and $\ev_y \colon X^Y \to X$ by $\ev_y(g) = g(y)$.
If $U$ and $V$ are $T$-algebras and $f \colon U \to V$ is a $T$-algebra homomorphism, then the image $\img(f)$ of $f$ is a $T$-algebra, with the $T$-algebra structure inherited from $V$.

The following proposition connects algebra homomorphisms from the free $T$-algebra on a set $U$ to an algebra $V$ with functions $U\to V$. We will make use of this later in the section.
\begin{proposition}
\label{prop:bijective}
	Given a set $U$ and a $T$-algebra $(V, v)$, there is a bijective correspondence between $T$-algebra homomorphisms $TU \to V$ and functions $U \to V$: for a $T$-algebra homomorphism $f \colon TU \to V$, define $f^\dagger = f \circ \eta \colon U \to V$; for a function $g \colon U \to V$, define $g^\sharp = v \circ Tg \colon TU \to V$.
Then $g^\sharp$ is a $T$-algebra homomorphism called the \emph{free $T$-extension of $g$}, and we have $f^{\dagger\sharp} = f$ and $g^{\sharp\dagger} = g$.
\end{proposition}
We now have all the ingredients to define our notion of automaton with side-effects and their language semantics.
We fix a monad $(T, \eta, \mu)$ with $T$ preserving finite sets, as well as a $T$-algebra $O$ that models outputs of automata.
%

\begin{definition}[$T$-automaton]
	A \emph{$T$-automaton} is a quadruple $(Q, \delta \colon Q \to Q^A, \out \colon Q \to O, \init \in Q)$, where the \emph{state space} $Q$ is a $T$-algebra, the \emph{transition map} $\delta$ and \emph{output map} $\out$ are $T$-algebra homomorphisms, and $\init$ is the \emph{initial state}.
\end{definition}
\begin{example}
	DFAs are $\mathtt{Id}$-automata when $O = 2 = \{0, 1\}$ is used to distinguish accepting from rejecting states.
	For the more general case of $O$ being any set, DFAs generalize into \emph{Moore automata}.
	\label{ex:dfa}
\end{example}
\begin{example}\label{ex:nfas}
Recall that $\Ps$-algebras are JSLs, and their homomorphisms are join-preserving functions.
In a $\Ps$-automaton, $Q$ is equipped with a join operation, and $Q^A$ is a join-semilattice with pointwise join: $(f \vee g)(a) = f(a) \vee g(a)$ for $a \in A$. Since the automaton maps preserve joins, we have, in particular, $\delta(q_1 \vee q_2)(a) = \delta(q_1)(a) \vee \delta(q_2)(a)$.
	One can represent an NFA over a set of states $S$ as a $\Ps$-automaton by taking $Q = (\Ps(S),\bigcup)$ and $O = 2$, the Boolean join-semilattice with the \emph{or} operation as its join.
	Let $\init \subseteq S$ be the set of initial states and $\out \colon \Ps(Q) \to 2$ and $\delta \colon \Ps(S) \to \Ps(S)^A$ the respective extensions (Proposition~\ref{prop:bijective}) of the NFA's output and transition functions.
	The resulting $\Ps$-automaton is precisely the determinized version of the NFA.
\end{example}
More generally, an automaton with side-effects given by a monad $T$ always represents a $T$-automaton with a free state space: by applying Proposition~\ref{prop:bijective}, we have the following.
\begin{proposition}\label{prop:succinct}
	A $T$-automaton of the form $((TX, \mu_X), \delta, \out, \init)$, for any set $X$, is completely defined by the set $X$ with the element $\init \in TX$ and functions
	\begin{align*}
		\delta^\dagger \colon X \to (TX)^A &
			&
			\out^\dagger \colon X \to O.
	\end{align*}
\end{proposition}
We call such a $T$-automaton a \emph{succinct} automaton, which we sometimes identify with the representation $(X, \delta^\dagger, \out^\dagger, \init)$.

%
%
A \emph{(generalized) language} is a function $\lang \colon A^* \to O$.
For every $T$-automaton we have an \emph{observability} and a \emph{reachability} map, telling respectively which state is reached by reading a given word and which language each state recognizes.
\begin{definition}[Reachability/observability maps]
The \emph{reachability map} of a $T$-automaton $\autom$ with state space $Q$ is a function $r_\autom \colon A^* \to Q$ inductively defined as follows:
	$r_\autom(\eword)
		= \init$
	and
	$r_\autom(ua)
		= \delta(r_\autom(u))(a)$.
The \emph{observability map} of $\autom$ is a function $o_\autom \colon Q \to O^{A^*}$ inductively defined as follows:
	$o_\autom(q)(\eword)
		= \out(q)$
	and
	$o_\autom(q)(av)
		= o_\autom(\delta(q)(a))(v)$.
\end{definition}
%
%
%
The \emph{language accepted by $\autom$} is the function $\lang_\autom \colon A^* \to O$ given by $\lang_\autom = o_\autom(\init) = \out_\autom \circ r_\autom$.
\begin{example}
For an NFA $\autom$ represented as a $\Ps$-automaton, as seen in Example~\ref{ex:nfas}, $o_\autom(q)$ is the language of $q$ in the traditional sense. Notice that $q$, in general, is a set of states: $o_\autom(q)$ takes the union of languages of singleton states. The set $\lang_\autom$ is the language accepted by the initial states, i.e., the language of the whole NFA. The reachability map $r_\autom(u)$ returns the set of states reached via all possible paths reading $u$.
\end{example}
Given a language $\lang \colon A^* \to O$, there exists a (unique) \emph{minimal $T$-automaton} $\minaut_\lang$ accepting $\lang$. Its existence follows from general facts see \citep[see, e.g.,][]{heerdt2016}.
%
\begin{definition}[Minimal $T$-automaton for $\lang$]\label{def:min-taut}
Let $t_\lang \colon A^* \to O^{A^*}$ be the function giving the \emph{residual languages} of $\lang$, namely $t_\lang(u) = \lambda v.\lang(uv)$.
The minimal $T$-automaton $\minaut_{\lang}$ accepting $\lang$ has state space $M = \img(t_\lang^\sharp)$, initial state $\init = t_\lang(\eword)$, and $T$-algebra homomorphisms $\out \colon M \to O$ and $\delta \colon M \to M^A$ given by $\out(t_\lang^\sharp(U)) = \lang(U)$ and $\delta(t_\lang^\sharp(U))(a)(v) = t_\lang^\sharp(U)(av)$.
\end{definition}
In the following, we will also make use of the \emph{minimal Moore automaton} accepting $\lang$. Although this always exists---it is defined by instantiating \definitionname~\ref{def:min-taut} with $T = \Id$---it need not be finite. The following property says that finiteness of Moore automata and of $T$-automata accepting the same language are intimately related.

\begin{restatable}{proposition}{finitet}
The minimal Moore automaton accepting $\lang$ is finite if and only if the minimal $T$-automaton accepting $\lang$ is finite.
\end{restatable}
%
%
%
%
%
%
%

\section{A General Algorithm}\label{sec:algorithm}

In this section we introduce our extension of $\lstar$ to learn automata with side-effects.
The algorithm is parametric in the notion of side-effect, represented as the monad $T$, and is therefore called $\lstarT$.
%
%
We fix a language $\lang \colon A^* \to O$ that is to be learned, and we assume that there is a finite $T$-automaton accepting $\lang$. This assumption generalizes the requirement of $\lstar$ that $\lang$ is regular (i.e., accepted by a specific class of $T$-automata, see \examplename~\ref{ex:dfa}).

%
%

An observation table consists of a pair of functions 
\[
	\Trow \colon S \to O^E \qquad\qquad \Brow \colon S \to (O^E)^A
\]
given by $\Trow(s)(e) = \lang(se)$ and $\Brow(s)(a)(e) = \lang(sae)$,
%
where $S, E \subseteq A^*$ are finite sets with $\eword \in S \cap E$. For $O=2$, we recover exactly the \lstar{} observation table. The key idea for \lstarT{} is defining closedness and consistency over the free $T$-extensions of those functions.
%
\begin{definition}[Closedness and Consistency]
	The table is \emph{closed} if for all $U \in T(S)$ and $a \in A$ there exists a $U' \in T(S)$ such that $\Trow^\sharp(U') = \Brow^\sharp(U)(a)$.
	The table is \emph{consistent} if for all $U_1, U_2 \in T(S)$ such that $\Trow^\sharp(U_1) = \Trow^\sharp(U_2)$ we have $\Brow^\sharp(U_1) = \Brow^\sharp(U_2)$.
	\label{def:clos-cons}
\end{definition}
For closedness, we do not need to check all elements of $T(S) \times A$ against elements of $T(S)$, but only those of $S \times A$, thanks to the following result.
\begin{lemma}
If for all $s \in S$ and $a \in A$ there is $U \in T(S)$ such that $\Trow^\sharp(U) = \Brow(s)(a)$, then the table is closed.
\label{lem:cls}
\end{lemma}
\begin{proof}
	Let $m \colon \img(\Trow^\sharp) \hookrightarrow O^E$ be the embedding of the image of $\Trow^\sharp$ into its codomain. 
	According to~\citet{CALF}, the definition of closedness given in Definition~\ref{def:clos-cons} amounts to requiring the existence of a $T$-algebra homomorphism $\mathsf{close}$ making the following diagram commute:
	\begin{equation}
		\begin{tikzcd}[ampersand replacement=\&]
			T(S) \ar[dashed]{d}[swap]{\mathsf{close}} \ar{dr}{\Brow^\sharp}  \\
			\img(\Trow^\sharp)^A \ar{r}[swap]{m^A} \& (O^E)^A
		\end{tikzcd}
	\label{diag:clos-emt}	
	\end{equation}
	It is easy to see that the hypothesis of this lemma corresponds to requiring the existence of a function $\mathsf{close}'$ making the diagram below on the left in $\Set$ commute.
	\begin{align*}
		\begin{gathered}
			\begin{tikzcd}[ampersand replacement=\&]
				S \ar[dashed]{d}[swap]{\mathsf{close}'} \ar{dr}{\Brow}  \\
				\img(\Trow^\sharp)^A \ar{r}[swap]{m^A} \& (O^E)^A
			\end{tikzcd}
		\end{gathered} &
			&
			\begin{gathered}
				\begin{tikzcd}[ampersand replacement=\&]
					T(S) \ar{d}[swap]{T(\mathsf{close}')} \ar{dr}{T(\Brow)}  \\
					T(\img(\Trow^\sharp)^A) \ar{r}[swap]{T(m^A)} \ar{d} \& T((O^E)^A) \ar{d} \\
					\img(\Trow^\sharp)^A \ar{r}[swap]{m^A} \& (O^E)^A
				\end{tikzcd}
			\end{gathered}
	\end{align*}
	This diagram can be made into a diagram of $T$-algebra homomorphisms as on the right, where the compositions of the left and right legs give respectively $\mathsf{close}'^\sharp$ and $\Brow^\sharp$. This diagram commutes because the top triangle commutes by functoriality of $T$, and the bottom square commutes by $m^A$ being a $T$-algebra homomorphism. Therefore we have that (\ref{diag:clos-emt}) commutes for $\mathsf{close} = \mathsf{close}'^\sharp$.
\end{proof}
\begin{example}\label{ex:tab}
For NFAs represented as $\Ps$-automata, the properties are as presented in Section~\ref{sec:overview-nfas}. Recall that for $T = \Ps$ and $O = 2$, the Boolean join-semilattice, $\Trow^\sharp$ and $\Brow^\sharp$ describe a table where rows are labeled by subsets of $S$. Then we have, for instance, $\Trow^\sharp(\{s_1,s_2\})(e) = \Trow(s_1)(e) \vee \Trow(s_2)(e)$, i.e., $\Trow^\sharp(\{s_1,s_2\})(e) = 1$ if and only if $\lang(s_1e) = 1$ or $\lang(s_2e) = 1$. Closedness amounts to check whether each row in the bottom part of the table is the join of a set of rows in the top part. Consistency amounts to check whether, for all sets of rows $U_1,U_2 \subseteq S$ in the top part of the table whose joins are equal, the joins of rows $U_1 \cdot \{a\}$ and $U_2 \cdot \{a\}$ in the bottom part are also equal, for all $a \in A$.
\end{example}
If closedness and consistency hold, we can define a hypothesis $T$-automaton $\hyp$. 
%
	Its state space is $H = \img(\Trow^\sharp)$, $\init = \Trow(\eword)$, and output and transition maps are given by:
	\begin{align*}
		\out &
			\colon H \to O &
			\out(\Trow^\sharp(U)) &
			= \Trow^\sharp(U)(\eword) \\
		\delta &
			\colon H \to H^A &
			\delta(\Trow^\sharp(U)) &
			= \Brow^\sharp(U).
	\end{align*}
The correctness of this definition follows from the abstract treatment of~\citet{CALF}, instantiated to the category of $T$-algebras and their homomorphisms.
%

\begin{figure}[t]
	\begin{codebox}
		\li $S, E \gets \{\eword\}$
		\li \Repeat
			\li \While the table is not closed or not consistent
				\li \Do \If the table is not closed
					\li \Then find $s \in S$, $a \in A$ such that $\Brow(s)(a) \neq \Trow^\sharp(U)$ for all $U \in T(S)$
					\li $S \gets S \cup \{sa\}$\label{line:clos-solve}
				\End
				\li \If the table is not consistent
					\li \Then find $U_1, U_2 \in T(S)$, $a \in A$, and $e \in E$ such that
					\zi\qquad $\Trow^\sharp(U_1) = \Trow^\sharp(U_2)$ and $\Brow^\sharp(U_1)(a)(e) \neq \Brow^\sharp(U_2)(a)(e)$
					\li $E \gets E \cup \{ae\}$
				\End
			\End
			\li\label{line:hyp} Construct the hypothesis $\hyp$ and submit it to the teacher
			\li\label{line:ce} \If the teacher replies \emph{no}, with a counterexample $z$
			\li \Then $S \gets S \cup \pref(z)$
		\End
		\li \Until the teacher replies \emph{yes}\label{line:eq-yes}
		\li \Return $\hyp$
	\end{codebox}
	\caption{Adaptation of \lstar{} for $T$-automata.}
	\label{fig:tlstar}
\end{figure}
We can now give our algorithm \lstarT{}.
In the same way as for the example in Section~\ref{sec:overview}, we only have to adjust lines \ref{line:closedness} and \ref{line:consistency} in \figurename~\ref{fig:lstar}.
The resulting algorithm is shown in \figurename~\ref{fig:tlstar}.

\paragraph{Correctness.}

Correctness for $\lstarT$ amounts to proving that, for any target language $\lang$, the algorithm terminates returning the minimal $T$-automaton $\minaut_\lang$ accepting $\lang$. As in the original $\lstar$ algorithm, we only need to prove that the algorithm terminates, that is, that only finitely many hypotheses are produced. Correctness follows from termination, since line~\ref{line:eq-yes} causes the algorithm to terminate only if the hypothesis automaton coincides with $\minaut_\lang$.

%
%
%
%
%
In order to show termination, we argue that the state space $H$ of the hypothesis increases while the algorithm loops, and that $H$ cannot be larger than $M$, the state space of $\minaut_\lang$. In fact, when a closedness defect is resolved (line~\ref{line:clos-solve}), a row that was not previously found in the image of $\Trow^\sharp \colon T(S) \to O^E$ is added, so the set $H$ grows larger.
When a consistency defect is resolved (line 9), two previously equal rows become distinguished, which also increases the size of $H$.

As for counterexamples, adding their prefixes to $S$ (line~\ref{line:ce}) creates a consistency defect, which will be fixed during the next iteration, causing $H$ to increase.
This is due to the following result, which says that the counterexample $z$ has a prefix that violates consistency.

\begin{proposition}\label{prop:correct}
	If $z \in A^*$ is such that $\lang_\hyp(z) \ne \lang(z)$ and $\pref(z) \subseteq S$, then there are a prefix $ua$ of $z$, with $u \in A^*$ and $a \in A$, and $U \in T(S)$ such that $\Trow(u) = \Trow^\sharp(U)$ and $\Brow(u)(a) \ne \Brow^\sharp(U)(a)$.
\end{proposition}
\begin{proof}
	Note that
	\begin{align*}
		\Trow(z)(\eword) &
			= \lang(z) &
			&
			\text{(definition of $\Trow$)} \\
		&
			\ne \lang_\hyp(z) &
			&
			\text{(assumption)} \\
		&
			= \out_\hyp(r_\hyp(z)) &
			&
			\text{(Definition of $\lang_\hyp$)} \\ 
		&
			= r_\hyp(z)(\eword) &
			&
			\text{(definition of $\out_\hyp$)},
	\end{align*}
	so $\Trow(z) \ne r_\hyp(z)$.
	Let $p \in A^*$ be the smallest prefix of $z$ satisfying $\Trow(p) \ne r_\hyp(p)$.
	We have $\Trow(\eword) = \init_\hyp = r_\hyp(\eword)$, so $p \ne \eword$ and therefore $p = ua$ for certain $u \in A^*$ and $a \in A$.
	Let $S' \subset S$ be the set from which $\hyp$ was constructed---recall that we added $\pref(z)$ to $S$ after constructing $\hyp$.
	Choose any $U \in T(S')$ such that $\Trow^\sharp(U) = r_\hyp(u)$, which is possible because $H$ is the image of $\Trow^\sharp$ restricted to the domain $T(S')$.
	By the minimality property of $p$ we have $\Trow(u) = r_\hyp(u) = \Trow^\sharp(U)$.
	Furthermore,
	\begin{align*}
		\Brow(u)(a) &
			= \Trow(ua) &
			&
			\text{(definitions of $\Trow$ and $\Brow$)} \\
		&
			\ne r_\hyp(ua) &
			&
			\text{($ua = p$ and $\Trow(p) \ne r_\hyp(p)$)} \\
		&
			= \delta_\hyp(r_\hyp(u))(a) &
			&
			\text{(definition of $r_\hyp$)} \\
		&
			= \delta_\hyp(\Trow^\sharp(U))(a) &
			&
			\text{($r_\hyp(u) = \Trow^\sharp(U)$)} \\
		&
			= \Brow^\sharp(U)(a) &
			&
			\text{(definition of $\delta_\hyp$)}.
		\qedhere
	\end{align*}
\end{proof}
%
%
Now, note that, by increasing $S$ or $E$, the hypothesis state space $H$ never decreases in size. Moreover, for $S = A^*$ and $E = A^*$, $\Trow^\sharp = t_\lang^\sharp$, as defined in Definition~\ref{cor:decomp}. Therefore, since $H$ and $M$ are defined as the images of $\Trow^\sharp$ and $t_\lang^\sharp$, respectively, the size of $H$ is bounded by that of $M$. Since $H$ increases while the algorithm loops, the algorithm must terminate and thus is correct. 

We note that the RFSA learning algorithm of Bollig et al.\ does not terminate using this counterexample processing method~\citep[Appendix~F]{Bollig09TR}.
This is due to their notion of consistency being weaker than ours: we have shown that progress is guaranteed because a consistency defect, in our sense, is created using this method.

\paragraph{Query complexity.}
The complexity of automata learning algorithms is usually measured in terms of the number of both membership and equivalence queries asked, as it is common to assume that computations within the algorithm are insignificant compared to evaluating the system under analysis in real-world applications. The complexity of answering the queries themselves is not considered, as it depends on the implementation of the teacher, which the algorithm abstracts from.


Notice that, as the table is a $T$-algebra homomorphism, asking membership queries for rows labeled by words in $S$ is enough to determine all other rows, for which queries need not be asked.
We measure the query complexities in terms of the number of states $n$ of the minimal Moore automaton, the number of states $t$ of the minimal $T$-automaton, the size $k$ of the alphabet, and the length $m$ of the longest counterexample.
Note that $t$ cannot be smaller than $n$, but it can be much bigger.
For example, when $T = \Ps$, $t$ may be in $\mathcal{O}(2^n)$.\footnote{%
	This can be seen from the language $\{a^p\}$, for some $p \in \N$ and a singleton alphabet $\{a\}$.
	Its residual languages are $\emptyset$ and $\{a^i\}$ for all $0 \le i \le p$, which means the minimal DFA accepting the language has $p + 2$ states.
	However, the residual languages w.r.t.\ sets of words are all the subsets of $\{\eword, a, aa, \ldots, a^p\}$---hence, the minimal $T$-automaton has $2^{p + 1}$ states.
}

The maximum number of closedness defects fixed by the algorithm is $n$, as a closedness defect for the setting with algebraic structure is also a closedness defect for the setting without that structure.
The maximum number of consistency defects fixed by the algorithm is $t$, as fixing a consistency defect distinguishes two rows that were previously identified.
Since counterexamples lead to consistency defects, this also means that the algorithm will not pose more than $t$ equivalence queries.
A word is added to $S$ when fixing a closedness defect, and $\mathcal{O}(m)$ words are added to $S$ when processing a counterexample.
The number of rows that we need to fill using queries is therefore in $\mathcal{O}(tmk)$.
The number of columns added to the table is given by the number of times a consistency defect is fixed and thus in $\mathcal{O}(t)$.
Altogether, the number of membership queries is in $\mathcal{O}(t^2mk)$.

\section{Succinct Hypotheses}\label{sec:hypotheses}

We now describe the first of two optimizations, 
which is enabled by the use of monads. Our algorithm produces hypotheses that can be quite large, as their state space is the image of $\Trow^\sharp$, which has the whole set $T(S)$ as its domain.
For instance, when $T = \Ps$, $T(S)$ is exponentially larger than $S$.
We show how we can compute \emph{succinct} hypotheses, whose state space is given by a subset of $S$. We start by defining sets of \emph{generators for the table}.
\begin{definition}
	A set $S' \subseteq S$ is a \emph{set of generators for the table} whenever for all $s \in S$ there is $U \in T(S')$ such that $\Trow(s) = \Trow^\sharp(U)$.\footnote{%
		Here and hereafter we assume that $T(S') \subseteq T(S)$, and more generally that $T$ preserves inclusion maps.
		To eliminate this assumption, one could take the inclusion map $f \colon S' \hookrightarrow S$ and write $\Trow^\sharp(T(f)(U))$ instead of $\Trow^\sharp(U)$.
	}
\end{definition}
%
Intuitively, $U$ is the decomposition of $s$ into a ``combination'' of generators. When $T = \Ps$, $S'$ generates the table whenever each row can be obtained as the join of a set of rows labeled by $S'$. Explicitly: for all $s \in S$ there is $\{s_1,\dots,s_n\} \subseteq S'$ such that $\Trow(s) = \Trow^\sharp(\{s_1,\dots,s_n\}) = \Trow(s_1) \vee \dots \vee \Trow(s_n)$.

Recall that $\hyp$, with state space $H$, is the hypothesis automaton for the table. The existence of generators $S'$ allows us to compute a $T$-automaton with state space $T(S')$ equivalent to $\hyp$. We call this the \emph{succinct hypothesis}, although $T(S')$ may be larger than $H$. Proposition~\ref{prop:succinct} tells us that the succinct hypothesis can be represented as an automaton with side-effects in $T$ that has $S'$ as its state space. This results in a lower space complexity when storing the hypothesis.

\newcommand{\TRowGen}{\widehat{\Trow}^\sharp}
We now show how the succinct hypothesis is computed. Observe that, if generators $S'$ exist, $\Trow^\sharp$ factors through the restriction of itself to $T(S')$. Denote this latter function $\TRowGen$. Since we have $T(S') \subseteq T(S)$, the image of $\TRowGen$ coincides with $\img(\Trow^\sharp) = H$, and
therefore the surjection restricting $\TRowGen$ to its image has the form $e \colon T(S') \to H$. Any right inverse $i \colon H \to T(S')$ of the function $e$ (that is, $e \circ i = \idf_H$, but whereas $e$ is a $T$-algebra homomorphism, $i$ need not be one) yields a succinct hypothesis as follows.
%
\begin{definition}[Succinct Hypothesis]
	The \emph{succinct hypothesis} is the following $T$-automaton $\suc$: its state space is $T(S')$, its initial state is $\init = i(\Trow(\eword))$, and we define
	\begin{align*}
		\out^\dagger &
			\colon S' \to O &
			\out^\dagger(s) &
			= \Trow(s)(\eword) \\
		\delta^\dagger &
			\colon S' \to T(S')^A &
			\delta^\dagger(s)(a) &
			= i(\Brow(s)(a)).
	\end{align*}
\end{definition}
This definition is inspired by that of a \emph{scoop}, due to~\citet{arbib1975_}.
%
%
\begin{restatable}{proposition}{succlang}
	Any succinct hypothesis of $\hyp$ accepts the same language as $\hyp$.
\end{restatable}
The proof can be found in the appendix. 
We now give a simple procedure to compute a \emph{minimal} set of generators, that is, a set $S'$ such that no proper subset is a set of generators.
This generalizes a procedure defined by~\citet{angluin2015} for non-deterministic, universal, and alternating automata.
\begin{restatable}{proposition}{genalg}\label{prop:gen-alg}
The following algorithm returns a minimal set of generators for the table:
\begin{codebox}
\zi $S' \gets S$
	\zi \While there are $s \in S'$ and $U \in T(S' \setminus \{s\})$ s.t.\ $\Trow^\sharp(U) = \Trow(s)$ \Do
\zi $S' \gets S' \setminus \{s\}$ \End
\zi \Return $S'$
\end{codebox}
\end{restatable}
%
%
%
The proof can be found in the appendix.
Determining whether $U$ as in the algorithm given in Proposition~\ref{prop:gen-alg} exists, one can always naively enumerate all possibilities, using that $T$ preserves finite sets.
This is what we call the basic algorithm.
For specific algebraic structures, one may find more efficient methods, as we show in the following example.
\begin{example}\label{ex:psefficient}
	Consider again the powerset monad $T = \Ps$. We now exemplify two ways of computing succinct hypotheses, which are inspired by the definitions of canonical RFSAs~\citep{denis2002}.
	The basic idea is to start from a deterministic automaton and to remove states that are equivalent to a set of other states.
	The algorithm given in Proposition~\ref{prop:gen-alg} computes a minimal $S'$ that only contains labels of rows that are not the join of other rows. (In case two rows are equal, only one of their labels is kept.) In other words, as mentioned in Section~\ref{sec:overview}, $S'$ contains labels of join-irreducible rows.
	To concretize the algorithm efficiently, we use a method introduced by~\citet{bollig2009}, which essentially exploits the natural order on the JSL of table rows.
	In contrast to the basic exponential algorithm, this results in a polynomial one.\footnote{When we refer to computational complexities, as opposed to query complexities, they are in terms of the sizes of $S$, $E$, and $A$.}
	Bollig et al.\ determine whether a row is a join of other rows by comparing the row just to the join of rows below it.
	Like them, we make use of this also to compute right inverses of $e$, for which we will formalize the order.

	The function $e \colon \Ps(S') \to H$ tells us which sets of rows are equivalent to a single state in $H$. We show two right inverses $H \to \Ps(S')$ for it. The first one,
	%
%
%
	\[
		i_1(h) = \{s \in S' \mid \Trow(s) \le h\},
	\]
	stems from the construction of the \emph{canonical RFSA} of a language~\citep{denis2002}.
	Here we use the order $a \le b \iff a \vee b = b$ induced by the JSL structure.
	The resulting construction of a succinct hypothesis was first used by~\citet{bollig2009}.
	This succinct hypothesis has a ``maximal'' transition function, meaning that no more transitions can be added without changing the language of the automaton.
	
	The second inverse is
	\begin{multline*}\label{eq:srfsa}
		i_2(h) = \{s \in S' \mid \Trow(s) \le h \text{ and for all } s' \in S' \text{ s.t. } \Trow(s) \le \Trow(s') \le h \\
		\text{ we have } \Trow(s) = \Trow(s')\},
	\end{multline*}
	resulting in a more economical transition function, where some redundancies are removed. This corresponds to the \emph{simplified canonical RFSA}~\citet{denis2002}.
\end{example}
%
%
\begin{example}
	Consider again the powerset monad $T = \Ps$, and recall the table in \figurename~\ref{fig:obs6}.
	When $S' = S$, the right inverse given by $i_1$ yields the succinct hypothesis shown below.
	\begin{center}
	\scalebox{.8}{\begin{automaton}
		\node[initial,state,accepting] (q0) {};
		\node[state] (q1) [right of=q0] {};
		\node[state,accepting] (q2) [right of=q1] {};
		\path[->]
		(q0) edge [bend left] node {$a$} (q1)
		(q1) edge [bend left] node {$a$} (q2)
		(q1) edge [bend left,swap] node {$a$} (q0)
		(q1) edge [loop above] node {$a$} ()
		(q2) edge [bend left,swap] node {$a$} (q1)
		(q2.south) edge [bend left,swap] node  [] {$a$} (q0.south)
		(q2) edge [loop right] node {$a$} ();
	\end{automaton}}
	\end{center}
	Note that $i_1(\Trow(aa)) = \{\eword, a, aa\}$. 
	When instead taking $i_2$, the succinct hypothesis is just the DFA (\ref{aut:ml}) because $i_2(\Trow(aa)) = \{aa\}$. Rather than constructing a succinct hypothesis directly, our algorithm first reduces the set $S'$.
	In this case, we note that $\Trow(aa) = \Trow^\sharp(\{\eword, a\})$, so we can remove $aa$ from $S'$.
	Now $i_1$ and $i_2$ coincide and produce the NFA (\ref{aut:n}).
	Minimizing the set $S'$ in this setting essentially comes down to determining what \citet{bollig2009} call the \emph{prime} rows of the table.
\end{example}
\begin{remark}
	The algorithm in Proposition~\ref{prop:gen-alg} implicitly assumes an order in which elements of $S$ are checked. Although the algorithm is correct for any such order, different orders may give results that differ in size.
\end{remark}


\newcommand{\h}{\mathcal{R}}

\section{Optimized Counterexample Handling}\label{sec:counterexamples}

The second optimization we give generalizes the 
%
counterexample processing method due to~\citet{rivest1993}, which improves the worst case complexity of the number of membership queries needed in \lstar{}.
\citet{maler1995} proposed to add all suffixes of the counterexample to the set $E$ instead of adding all prefixes to the set $S$.
This eliminates the need for consistency checks in the deterministic setting.
The method by Rivest and Schapire finds a \emph{single} suffix of the counterexample and adds it to $E$.
This suffix is chosen in such a way that it either distinguishes two existing rows or creates a closedness defect, both of which imply that the hypothesis automaton will grow.

%


The main idea is finding the distinguishing suffix via the hypothesis automaton $\hyp$.
Given a word $u \in A^*$, let $q_u$ be the state in $\hyp$ reached by reading $u$, i.e., $q_u = r_\hyp(u)$.
For each $q \in H$, we pick any $U_q \in T(S)$ that yields $q$ according to the table, i.e., such that $\Trow^\sharp(U_q) = q$.
Then for a counterexample $z$ we have that the residual language w.r.t.\ $U_{q_z}$ does not ``agree'' with the residual language w.r.t.\ $z$.

The above intuition can be formalized as follows.
Let $\h \colon A^* \to O^{A^*}$ be given by $\h(u) = t_\lang^\sharp(U_{q_u})$ for all $u \in A^*$, the residual language computation. 
We have the following technical lemma, saying that a counterexample $z$ distinguishes the residual languages $t_\lang(z)$ and $\h(z)$.
%
%
%

\begin{lemma}\label{lem:ce}
	If $z \in A^*$ is such that $\lang_\hyp(z) \ne \lang(z)$, then $t_\lang(z)(\eword) \ne \h(z)(\eword)$.
\end{lemma}
\begin{proof}
	We have
	\begin{align*}
		t_\lang(z)(\eword) &
			= \lang(z) &
			&
			\text{(definition of $t_\lang$)} \\
		&
			\ne \lang_\hyp(z) &
			&
			\text{(assumption)} \\
		&
			= (\out_\hyp \circ r_\hyp)(z) &
			&
			\text{(definition of $\lang_\hyp$)} \\
		&
			= r_\hyp(z)(\eword) &
			&
			\text{(definition of $\out_\hyp$)} \\
		&
			= q_z(\eword) &
			&
			\text{(definition of $q_z$)} \\
		&
			= \Trow^\sharp(U_{q_z})(\eword) &
			&
			\text{(definition of $U_{q_z}$)} \\
		&
			= t_\lang^\sharp(U_{q_z})(\eword) &
			&
			\text{(definitions of $\Trow$ and $t_\lang$)} \\
		&
			= \h(z)(\eword) &
			&
			\text{(definition of $\h$)}.
		\qedhere
	\end{align*}
\end{proof}
We assume that $U_{q_\eword} = \eta(\eword)$.
For a counterexample $z$, we then have $\h(\eword)(z) = t_\lang(\eword)(z) \ne \h(z)(\eword)$. While reading $z$, the hypothesis automaton passes a sequence of states $q_{u_0}$, $q_{u_1}$,$q_{u_2}$,\dots,$q_{u_n}$, where $u_0 = \epsilon$, $u_n = z$, and $u_{i+1} = u_ia$ for some $a \in A$ is a prefix of $z$. If $z$ were correctly classified by $\hyp$, all residuals $\h(u_i)$ would classify the remaining suffix $v$ of $z$, i.e., such that $z = u_iv$, in the same way. However, the previous lemma tells us that, for a counterexample $z$, this is not case, meaning that for some suffix $v$ we have $\h(ua)(v) \neq \h(u)(av)$. In short, this inequality is discovered along a transition in the path to $z$. 
%
%
%
%
%
\begin{corollary}\label{cor:decomp}
	If $z \in A^*$ is such that $\lang_\hyp(z) \ne \lang(z)$, then 
	there are $u, v \in A^*$ and $a \in A$ such that $uav = z$ and $\h(ua)(v) \ne \h(u)(av)$.
\end{corollary}
%
To find such a decomposition efficiently, Rivest and Schapire use a binary search algorithm.
We conclude with the following result that turns the above property into the elimination of a closedness witness.
That is, given a counterexample $z$ and the resulting decomposition $uav$ from the above corollary, we show that, while currently $\Trow^\sharp(U_{q_{ua}}) = \Brow^\sharp(U_{q_u})(a)$, after adding $v$ to $E$ we have $\Trow^\sharp(U_{q_{ua}})(v) \ne \Brow^\sharp(U_{q_u})(a)(v)$.
(To see that the latter follows from the proposition below, note that for all $U \in T(S)$ and $e \in E$, $\Trow^\sharp(U)(e) = t_\lang^\sharp(U)(e)$ and for each $a' \in A$, $\Brow^\sharp(U)(a')(e) = t_\lang^\sharp(U)(a'e)$, by the definition of those maps.)
The inequality means that either we have a closedness defect, or there still exists some $U \in T(S)$ such that $\Trow^\sharp(U) = \Brow^\sharp(U_{q_u})(a)$.
In this case, the rows $\Trow^\sharp(U)$ and $\Trow^\sharp(U_{q_{ua}})$ have become distinguished by adding $v$, which means that the size of $H$ has been increased.
We know that a closedness defect leads to an increase in the size of $H$, so in any case we make progress.
\begin{proposition}\label{prop:unclosed}
	If $z \in A^*$ is such that $\lang_\hyp(z) \ne \lang(z)$, then there are $u, v \in A^*$ and $a \in A$ such that $\Trow^\sharp(U_{q_{ua}}) = \Brow^\sharp(U_{q_u})(a)$ and $t_\lang^\sharp(U_{q_{ua}})(v) \ne t_\lang^\sharp(U_{q_u})(av)$.
\end{proposition}
\begin{proof}
	By Corollary~\ref{cor:decomp} we have $u, v \in A^*$ and $a \in A$ such that $\h(ua)(v) \ne \h(u)(av)$.
	This directly yields the inequality by the definition of $\h$.
	Furthermore,
	\begin{align*}
		\Trow(U_{q_{ua}}) &
			= q_{ua} &
			&
			\text{(definition of $U_{q_{ua}}$)} \\
		&
			= r_\hyp(ua) &
			&
			\text{(definition of $q_{ua}$)} \\
		&
			= \delta_\hyp(r_\hyp(u))(a) &
			&
			\text{(definition of $r_\hyp$)} \\
		&
			= \delta_\hyp(q_u)(a) &
			&
			\text{(definition of $q_u$)} \\
		&
			= \delta_\hyp(\Trow^\sharp(U_{q_u}))(a) &
			&
			\text{(definition of $U_{q_u}$)} \\
		&
			= \Brow^\sharp(U_{q_u})(a) &
			&
			\text{(definition of $\delta_\hyp$)}.
		\qedhere
	\end{align*}
\end{proof}
We now show how to combine this optimized counterexample processing method with the succinct hypothesis optimization from Section~\ref{sec:hypotheses}.
Recall that the succinct hypothesis $\suc$ is based on a right inverse $i \colon H \to T(S')$ of $e \colon T(S') \to H$.
Choosing such an $i$ is equivalent to choosing $U_q$ for each $q \in H$.
We then redefine $\h$ using the reachability map of the succinct hypothesis.
Specifically, $\h(u) = t_\lang^\sharp(r_\suc(u))$ for all $u \in A^*$.

Unfortunately, there is one complication.
We assumed earlier that $U_{q_\eword} = \eta(\eword)$, or more specifically $\h(\eword)(z) = \lang(z)$. This now may be impossible because we do not even necessarily have $\eword \in S'$.
We show next that if this equality does not hold, then there are two rows that we can distinguish by adding $z$ to $E$.
Thus, after testing whether $\h(\eword)(z) = \lang(z)$, we either add $z$ to $E$ (if the test fails) or proceed with the original method.
\begin{proposition}\label{prop:uninit}
	If $z \in A^*$ is such that $\h(\eword)(z) \ne \lang(z)$, then $\Trow^\sharp(\init_\suc) = \Trow(\eword)$ and $t_\lang^\sharp(\init_\suc)(z) \ne t_\lang(\eword)(z)$.
\end{proposition}
\begin{proof}
	We have $\Trow^\sharp(\init_\suc) = \Trow^\sharp(i(\Trow(\eword))) = \Trow(\eword)$ by the definitions of $\init_\suc$ and $i$, and
	\begin{align*}
		t_\lang^\sharp(i(\Trow(\eword)))(z) &
			= t_\lang^\sharp(\init_\suc)(z) &
			&
			\text{(definition of $\init_\suc$)} \\
		&
			= t_\lang^\sharp(r_\suc(\eword))(z) &
			&
			\text{(definition of $r_\suc$)} \\
		&
			= \h(\eword)(z) &
			&
			\text{(definition of $\h$)} \\
		&
			\ne \lang(z) &
			&
			\text{(assumption)} \\
		&
			= t_\lang(\eword)(z) &
			&
			\text{(definition of $t_\lang$)}.
		\qedhere
	\end{align*}
\end{proof}
To see that the original method still works, we prove the analogue of Lemma~\ref{lem:ce} for the new definition of $\h$.
\begin{lemma}
	If $z \in A^*$ is such that $\lang_\suc(z) \ne \lang(z)$ and $\h(\eword)(z) = \lang(z)$, then $\h(\eword)(z) \ne \h(z)(\eword)$.
\end{lemma}
\begin{proof}
	We have
	\begin{align*}
		\h(\eword)(z) &
			= \lang(z) &
			&
			\text{(assumption)} \\
		&
			\ne \lang_\suc(z) &
			&
			\text{(counterexample)} \\
		&
			= (\out_\suc \circ r_\suc^\dagger)(z) &
			&
			\text{(definition of $\lang_\suc$)} \\
		&
			= (\Trow^\sharp \circ r_\suc^\dagger)(z)(\eword) &
			&
			\text{(definition of $\out_\suc$)} \\
		&
			= t_\lang^\sharp(r_\suc^\dagger(z))(\eword) &
			&
			\text{(definition of $\Trow^\sharp$)} \\
		&
			= \h(z)(\eword) &
			&
			\text{(definition of $\h$)}.
			\qedhere
	\end{align*}
\end{proof}
\begin{corollary}\label{cor:decompsuc}
	If $z \in A^*$ is such that $\lang_\suc(z) \ne \lang(z)$ and $\h(\eword)(z) = \lang(z)$, then there are $u, v \in A^*$ and $a \in A$ such that $uav = z$ and $\h(ua)(v) \ne \h(u)(av)$.
\end{corollary}
Now we are ready to prove the analogue of Proposition~\ref{prop:unclosed}.
\begin{proposition}
	If $z \in A^*$ is such that $\lang_\suc(z) \ne \lang(z)$ and $\h(\eword)(z) = \lang(z)$, then there are $u, v \in A^*$ and $a \in A$ such that $\Trow^\sharp(r_\suc^\dagger(ua)) = \Brow^\sharp(r_\suc^\dagger(u))(a)$ and $t_\lang^\sharp(r_\suc^\dagger(ua))(v) \ne t_\lang^\sharp(r_\suc^\dagger(u))(av)$.
\end{proposition}
\begin{proof}
	Let $u$, $a$, and $v$ be as in Corollary~\ref{cor:decompsuc}.
	Thus,
	\[
		t_\lang^\sharp(r_\suc^\dagger(ua))(v) = \h(ua)(v) \ne \h(u)(av) = t_\lang^\sharp(r_\suc^\dagger(u))(av).
	\]
	Furthermore, since for all $s \in S$ and $b \in A$ we have
	\begin{align*}
		((\Trow^\sharp)^A \circ \delta_\suc^\dagger)(s)(b) &
			= \Trow^\sharp(\delta_\suc^\dagger(s)(b)) \\
		&
			= (\Trow^\sharp \circ i)(\Brow(s)(b)) &
			&
			\text{(definition of $\delta_\suc^\dagger$)} \\
		&
			= \Brow(s)(b) &
			&
			\text{(definition of $i$)},
	\end{align*}
	it follows that $(\Trow^\sharp)^A \circ \delta_\suc = \Brow^\sharp$.
	Therefore,
	\begin{align*}
		\Trow^\sharp(r_\suc^\dagger(ua)) &
			= \Trow^\sharp(\delta_\suc(r_\suc^\dagger(u))(a)) &
			&
			\text{(definition of $r_\suc^\dagger$)} \\
		&
			= ((\Trow^\sharp)^A \circ \delta_\suc)(r_\suc^\dagger(u))(a) \\
		&
			= \Brow^\sharp(r_\suc^\dagger(u))(a).
			&
			&
			\qedhere
	\end{align*}
\end{proof}
\begin{example}
	Recall the succinct hypothesis $\suc$ from \figurename~\ref{fig:hyp2} for the table in \figurename~\ref{fig:obs1}.
	Note that $S' = S$ cannot be further reduced.
	The hypothesis is based on the right inverse $i \colon H \to \Ps(S)$ of $e \colon \Ps(S) \to H$ given by $i(\Trow(\eword)) = \{\eword\}$ and $i(\Trow^\sharp(\emptyset)) = \emptyset$.
	This is the only possible right inverse because $e$ is bijective.
	For the prefixes of the counterexample $aa$ we have $r_\suc(\eword) = \{\eword\}$ and $r_\suc(a) = r_\suc(aa) = \emptyset$.
	Note that $t_\lang^\sharp(\{\eword\})(aa) = 1$ while $t_\lang(\emptyset)(a) = t_\lang(\emptyset)(\eword) = 0$.
	Thus, $\h(\eword)(aa) \ne \h(a)(a)$.
	Adding $a$ to $E$ would indeed create a closedness defect.
\end{example}
\paragraph{Query complexity.}
Again, we measure the membership and equivalence query complexities in terms of the number of states $n$ of the minimal Moore automaton, the number of states $t$ of the minimal $T$-automaton, the size $k$ of the alphabet, and the length $m$ of the longest counterexample.

A counterexample now gives an additional column instead of a set of rows, and we have seen that this leads to either a closedness defect or to two rows being distinguished.
Thus, the number of equivalence queries is still at most $t$, and the number of columns is still in $\mathcal{O}(t)$.
However, the number of rows that we need to fill using membership queries is now in $\mathcal{O}(nk)$.
This means that a total of $\mathcal{O}(tnk)$ membership queries is needed to fill the table.

Apart from filling the table, we also need queries to analyze counterexamples.
The binary search algorithm mentioned after Corollary~\ref{cor:decomp} requires for each counterexample $\mathcal{O}(\log{m})$ computations of $\h(x)(y)$ for varying words $x$ and $y$.
Let $r$ be the maximum number of queries required for a single such computation.
Note that for $u, v \in A^*$, and letting $\alpha \colon TO \to O$ be the algebra structure on $O$, we have
\[
	\h(u)(v) = \alpha(T(\ev_v \circ t_\lang)(U_{q_u}))
\]
for the original definition of $\h$ and
\[
	\h(u)(v) = \alpha(T(\ev_v \circ t_\lang)(r_\suc^\dagger(u)))
\]
in the succinct hypothesis case.
Since the restricted map $T(\ev_v \circ t_\lang) \colon TS \to TO$ is completely determined by $\ev_v \circ t_\lang \colon S \to O$, $r$ is at most $|S|$, which is bounded by $n$ in this optimized algorithm.
For some examples (see for instance the writer automata in Section~\ref{sec:examples}), we even have $r = 1$.
The overall membership query complexity is $\mathcal{O}(tnk + tr\log{m})$.

\paragraph{Dropping Consistency.}
We described the counterexample processing method based around Proposition~\ref{prop:unclosed} in terms of the succinct hypothesis $\suc$ rather than the actual hypothesis $\hyp$ by showing that $\h$ can be defined using $\suc$.
Since the definition of the succinct hypothesis does not rely on the property of consistency to be well-defined, this means we could drop the consistency check from the algorithm altogether.
We can still measure progress in terms of the size of the set $H$, but it will not be the state space of an actual hypothesis during intermediate stages.
This observation also explains why~\citet{bollig2009} are able to use a weaker notion of consistency in their algorithm. 
Interestingly, they exploit the canonicity of their choice of succinct hypotheses to arrive at a polynomial membership query complexity that does not involve the factor $t$.

\section{Examples}\label{sec:examples}

In this section we list several examples that can be seen as $T$-automata and hence learned via an instance of $\lstarT$. We remark that, since our algorithm operates on finite structures (recall that $T$ preserves finite sets), for each automaton type one can obtain a basic, correct-by-construction instance of $\lstarT$ for free, by just plugging the concrete definition of the monad into the abstract algorithm.
%
%
However, we note that this is not how $\lstarT$ is intended to be used in a real-world context. Instead, it should be seen as an abstract specification of the operations each concrete implementation needs to perform, or, in other words, as a template for real implementations.


%
For each instance below, we discuss whether certain operations admit a more efficient implementation than the basic one, based on the specific algebraic structure induced by the monad. We also mention related algorithms from the literature.
%
%
%
Due to our general treatment, the optimizations of Sections~\ref{sec:hypotheses}~and~\ref{sec:counterexamples} apply to all of these instances. 

%
%
\paragraph{Non-deterministic automata.}
As discussed before, non-deterministic automata are $\Ps$-automata with a free state space, provided that $O = 2$ is equipped with the ``or'' operation as its $\Ps$-algebra structure. 
We also mentioned that, as~\citet{bollig2009} showed, there is a polynomial time algorithm to check whether a given row is the join of other rows. This gives an efficient method for handling closedness straight away.
Moreover, as shown in \examplename~\ref{ex:psefficient}, it allows for an efficient construction of the succinct hypothesis.
Unfortunately, checking for consistency defects seems to require a number of computations exponential in the number of rows. We recall that~\citet{bollig2009} use an ad-hoc version of consistency which cannot be easily captured in our framework. However, as explained at the end of Section~\ref{sec:counterexamples}, we can in fact drop consistency altogether.

%
\paragraph{Universal automata.}
Just like non-deterministic automata, universal automata can be seen as $\Ps$-automata with a free state space.
The difference, however, is that the $\Ps$-algebra structure on $O = 2$ is dual: it is given by the ``and'' rather than the ``or'' operation.
Universal automata accept a word when all paths reading that word are accepting.
One can dualize the optimized specific algorithms for the case of non-deterministic automata.
This is precisely what~\citet{angluin2015} have done.

\paragraph{Partial automata.}
\newcommand{\May}{\mathtt{Maybe}}
Consider the \emph{maybe monad} $\May$, given by $\May(X) = 1 + X$, with natural transformations having components $\eta_X \colon X \to 1 + X$ and $\mu_X \colon 1 + 1 + X \to 1 + X$ defined in the standard way.
Partial automata with states $X$ can be represented as $\May$-automata with state space $\May(X) = 1 + X$, where there is an additional \emph{sink state}, and output algebra $O = \May(1) = 1 + 1$.
Here the left value is for rejecting states, including the sink one. The transition map $\delta \colon 1 + X \to (1 + X)^A$ represents an undefined transition as one going to the sink state.
The algorithm $\lstarT[\May]$ is mostly like \lstar{}, except that implicitly the table has an additional row with zeroes in every column.
Since the monad only adds a single element to each set, there is no need to optimize the basic algorithm for this specific case.
\paragraph{Weighted automata.}
Recall from Section~\ref{sec:prelim} the \emph{free semimodule monad} $\VMon$, sending a set $X$ to the free semimodule over a finite semiring $\sring$.
Weighted automata over a set of states $X$ can be represented as $\VMon$-automata whose state space is the semimodule $\VMon(X)$, the output function $\out \colon \VMon(X) \to \sring$ assigns a weight to each state, and the transition map $\delta \colon \VMon(X) \to \VMon(X)^A$ sends each state and each input symbol to a linear combination of states.
The obvious semimodule structure on $\sring$ extends to a pointwise structure on the potential rows of the table.
The basic algorithm loops over all linear combinations of rows to check closedness and over all pairs of combinations of rows to check consistency.
This is an extremely expensive operation.
If $\sring$ is a field, a row can be decomposed into a linear combination of other rows in polynomial time using standard techniques from linear algebra.
As a result, there are efficient procedures for checking closedness and constructing succinct hypotheses.
It was shown by~\citet{CALF} that consistency in this setting is equivalent to closedness of the transpose of the table.
This trick is due to~\citet{bergadano1996}, who first studied learning of weighted automata.
\paragraph{Alternating automata.}
\newcommand{\PPMon}{\mathtt{A}}
\newcommand{\upclos}{\uparrow\!\!}
We use the characterization of alternating automata due to~\citet{Meven}.
Recall that, given a partially ordered set $(P,\leq)$, an \emph{upset} is a subset $U$ of $P$ such that, if $x \in U$ and $x \leq y$, then $y \in U$. Given $Q \subseteq P$, we write $\upclos Q$ for the \emph{upward closure} of $Q$, that is the smallest upset of $P$ containing $Q$.
We consider the monad $\PPMon$ that maps a set $X$ to the set of all upsets of $\Ps(X)$.
Its unit is given by $\eta_X(x) = \upclos \{\{ x \}\}$ and its multiplication by
\[
	\mu_X(U) = \{V \subseteq X \mid \exists_{W \in U}\,\forall_{Y \in W}\,\exists_{Z \in Y}\,Z \subseteq V\}.
\]
The sets of sets in $\PPMon(X)$ can be seen as DNF formulae over elements of $X$, where the outer powerset is disjunctive and the inner one is conjunctive.
Accordingly, we define an algebra structure $\beta \colon \PPMon(2) \to 2$ on the output set $2$ by letting $\beta(U) = 1$ if $\{ 1 \} \in U$, 0 otherwise.
Alternating automata with states $X$ can be represented as $\PPMon$-automata with state space $\PPMon(X)$, output map $\out \colon \PPMon(X) \to 2$, and transition map $\delta \colon \PPMon(X) \to \PPMon(X)^A$, sending each state to a DNF formula over $X$.
The only difference with the usual definition of alternating automata is that $\PPMon(X)$ is not the full set $\Ps\Ps(X)$, which would not give a monad in the desired way.
However, for each formula in $\Ps\Ps(X)$ there is an equivalent one in $\PPMon(X)$.

An adaptation of \lstar{} for alternating automata was introduced by~\citet{angluin2015} and further investigated by~\citet{BerndtLLR17}. The former found that given a row $r \in 2^E$ and a set of rows $X \subseteq 2^E$, $r$ is equal to a DNF combination of rows from $X$ (where logical operators are applied component-wise) if and only if it is equal to the combination defined by
\[
	Y = \{\{x \in X \mid x(e) = 1\} \mid e \in E \wedge r(e) = 1\}.
\]
In our setting, we can reuse this idea to efficiently find closedness defects and to construct the hypothesis. Notice that, even though the monad $\PPMon$ formally requires the use of DNF formulae representing upsets, in the actual implementation we can use smaller formulae, e.g., $Y$ above instead of its upward closure. In fact, it is easy to check that DNF combinations of rows are invariant under upward closure.
%
%
%
As with non-deterministic and universal automata, we do not know of an efficient way to ensure consistency. As in the existing algorithms mentioned above, we could drop it altogether.
\paragraph{Writer automata.}
The examples considered so far involve existing classes of automata. To further demonstrate the generality of our approach, we introduce a new (as far as we know) type of automaton, which we call \emph{writer automaton}.

The \emph{writer monad} $\MMon(X) = \mon \times X$ for a finite monoid $\mon$ has a unit $\eta_X \colon X \to \mon \times X$ given by adding the unit $e$ of the monoid, $\eta_X(x) = (e, x)$, and a multiplication $\mu_X \colon \mon \times \mon \times X \to \mon \times X$ given by performing the monoid multiplication, $\mu_X(m_1, m_2, x) = (m_1m_2, x)$.
In Haskell, the writer monad is used for such tasks as collecting successive log messages, where the monoid is given by the set of sets or lists of possible messages and the multiplication adds a message.

The algebras for this monad are sets $Q$ equipped with an $\mon$-action. One may take the output object to be the set $\mon$ with the monoid multiplication as its action.
$\MMon$-automata with a free state space can be represented as deterministic automata that have an element of $\mon$ associated with each transition.
The semantics of these is that the encountered $\mon$-elements multiply along paths and finally multiply with the output of the last state to produce the actual output.

The basic learning algorithm is already of polynomial time complexity. In fact, to determine whether a given row is a combination of rows in the table, i.e., whether it is given by a monoid value applied to one of the rows in the table, one simply tries all of these values.
%
This allows us to check for closedness, to minimize the generators, and to construct the succinct hypothesis, in polynominal time.
Consistency involves comparing all ways of applying monoid values to rows and, for each comparison, at most $|A|$ further comparisons between one-letter extensions. The total number of comparisons is clearly polynomial in $|\mon|$, $|S|$ and $|A|$.

\section{Implementation}\label{sec:implementation}

We have implemented the general \lstarT{} algorithm in Haskell, taking full advantage of the monads provided by its standard library.
Apart from the high-level implementation, our library provides
\begin{itemize}
	\item
		a basic implementation for weighted automata over a finite semiring, with a polynomial time variation for the case where the semiring is a field\footnote{
			Despite the assumption in the present paper that the monad preserves finite set, our implementation can learn weighted automata over infinite fields and thus implements the general algorithm introduced by~\citet{bergadano1996}, which was studied in a categorical context by~\citet{jacobs2014} and~\citet{CALF}.
		};
	\item
		an implementation for non-deterministic automata that has polynomial time implementations for ensuring closedness and constructing the hypothesis, but not for ensuring consistency;
	\item
		a variation on the previous algorithm that uses the notion of consistency defined by~\citet{bollig2009};
	\item
		instantiations of the basic algorithm to the monad being $(-) + E$, for $E$ a finite set of exceptions, and $\MMon$, both of which result in polynomial time algorithms;
\end{itemize}
In this section we describe the main structure and ingredients of our library.
After recalling monads in Haskell in Section~\ref{sec:monads}, we start with the formalization of automata in Section~\ref{sec:imp-aut}.
We then introduce teachers in Section~\ref{sec:imp-teach} before exploring the actual learning algorithm in Section~\ref{sec:imp-learn}.
We give details for the non-deterministic and weighted case, whose monads deserve a closer analysis. 

\subsection{Monads}\label{sec:monads}

We note that a monad in Haskell is specified as a \emph{Kleisli triple} $(T, \eta, (-)^\sharp)$, where $T$ assigns to every set $X$ a set $TX$, $\eta$ consists of a component $\eta_X \colon X \to TX$ for each set $X$, and $(-)^\sharp$ provides for each function $f \colon X \to TY$ an extension $f^\sharp \colon TX \to TY$.
These need to satisfy
\begin{align*}
	f^\sharp \circ \eta &
		= f &
		\eta^\sharp &
		= \idf &
		(g^\sharp \circ f)^\sharp &
		= g^\sharp \circ f^\sharp.
\end{align*}
Kleisli triples are in a one-to-one correspondence with monads.
On both sides of this correspondence we have the same $T$ and $\eta$, which for a Kleisli triple are turned into a functor with a natural transformation by setting $Tf = (\eta \circ f)^\sharp$.
Furthermore, $(-)^\sharp$ and $\mu$ are obtained from each other by $f^\sharp = \mu \circ Tf$ and $\mu = \idf^\sharp$.
Indeed, under this correspondence the $(-)^\sharp$ operation is a specific instance of the extension operation defined for a monad, with the $T$-algebra codomain restricted to free $T$-algebras.
In Haskell, the $\eta$ of the Kleisli triple is written \textsf{return}, and, given $f \colon X \to TY$ and $x \in TX$, $f^\sharp(x)$ is written \texttt{x >>= f} and referred to as the \emph{bind} operation.
Furthermore, for any $f \colon X \to Y$, $Tf$ is given by \texttt{fmap f}.

Some basic $\Set$ monads cannot directly be written down in Haskell because their definition can only be given on types equipped with an equality check, or, for reasons of efficiency, a total order.
For example, the \texttt{Set} type provided by \texttt{Data.Set} comes with a \texttt{union} function that has the following signature:
\begin{verbatim}
union :: Ord a => Set a -> Set a -> Set a
\end{verbatim}
One will have to use unions in one way or another in defining the bind of the powerset monad.
However, since this bind needs to be of type
\begin{verbatim}
(>>=) :: Set a -> (a -> Set b) -> Set b
\end{verbatim}
and does not assume an \texttt{Ord} instance on \texttt{b}, the powerset monad cannot be defined in this way.

One solution is to delay the monadic computations in a wrapper type whose constructors are used to define a monad instance: the free monad.
Specifically, we endow the \emph{freer monad} of~\citet{kiselyov2015} with a constraint parameter:
\begin{verbatim}
data CFree c m a where
    Return :: a -> CFree c m a
    Bind :: (c b) => m b -> (b -> CFree c m a) -> CFree c m a
\end{verbatim}
Such a constrained free monad was first defined by George Giorgidze, but only for the specific case where \texttt{m} is \texttt{Set} and \texttt{c} is \texttt{Ord}.\footnote{\url{https://hackage.haskell.org/package/set-monad-0.2.0.0}}
On the constrained free monad we can define a complete \texttt{Monad} instance:
\begin{verbatim}
instance Monad (CFree c m) where
    return = Return
    f >>= g = case f of
        Return a -> g a
        Bind s h -> s `Bind` (h >=> g)
\end{verbatim}
This is the same code as used by~\citet{kiselyov2015}, but we note that on the last line, since \texttt{s} is the first argument of \texttt{Bind} in \texttt{f}, we know that the appropriate constrained needed to invoke \texttt{Bind} on the right-hand side, with again \texttt{s} as its first argument, is satisfied.

Finally, if there is a monad that is defined only on types satisfying a certain constraint, then we can convert from our free monad type with that constraint back to the actual ``monad'':
\begin{verbatim}
class ConstrainedMonad c m | m -> c where
    constrainedReturn :: (c a) => a -> m a
    constrainedBind :: (c a, c b) => m b -> (b -> m a) -> m a

unCFree :: (ConstrainedMonad c m, c a) => CFree c m a -> m a
unCFree f = case f of
    Return a -> constrainedReturn a
    Bind s g -> s `constrainedBind` (unCFree . g)
\end{verbatim}
Note that operations such as equality checks for \texttt{CFree c m} use \texttt{unCFree} to delegate the operation to whatever is defined for \texttt{m}.
This means that in code that abstracts from the monad we seem to be working with \texttt{m} as a monad.

As an example, the \texttt{Set} ``monad'' becomes
\begin{verbatim}
instance ConstrainedMonad Ord Set where
    constrainedReturn = Set.singleton
    s `constrainedBind` f = Set.unions [f a | a <- Set.toList s]
\end{verbatim}
We may then use \texttt{CFree Ord Set} as the monad.

To implement the free semimodule monad in Haskell, we use the \texttt{Map} type from \texttt{Data.Map}.
Note that the monad will be defined in the first argument for that type, so we need to create an auxiliary type to swap the arguments.
\begin{verbatim}
newtype Linear s k = Linear {fromLinear :: Map k s}
\end{verbatim}
Defining the monad again requires \texttt{Ord} constraints.
\begin{verbatim}
instance (Semiring s, Eq s) => ConstrainedMonad Ord (Linear s) where
    constrainedReturn a = Linear $ Map.singleton a mempty
    l `constrainedBind` f = Linear .
        foldl' (\m (k, s) -> ladd m . lscale s . fromLinear $ f k) Map.empty .
        Map.toList . lminimize $ fromLinear l
\end{verbatim}
The function \texttt{lscale} scales a map by an element from the semiring; \texttt{ladd} adds two maps together.
Both operations are pointwise.
The monad we can use is \texttt{CFree Ord (Linear s)}.

\subsection{Automata}\label{sec:imp-aut}

We model an automaton as a simple deterministic automaton.
\begin{verbatim}
data Aut a o q = Aut {
    initial :: q,
    delta :: q -> a -> q,
    out :: q -> o }
\end{verbatim}
For such automata, we can easily implement reachability and language functions, as well as bisimulation.
Bisimulation is used to realize exact equivalence queries for the teachers that hold an automaton accepting the language to be learned.
To optimize for the monad in the same way the learning algorithm is optimized, we use \emph{bisimulation up to context}~\citep{sangiorgi1998,rot2013}.
\begin{verbatim}
bisimT :: (Eq o) => ((t q, t r) -> [(t q, t r)] -> Bool) ->
    [a] -> Aut a o (t q) -> Aut a o (t r) -> Maybe [a]
\end{verbatim}
Here \texttt{t} represents the monad that we optimize for.
Up to context means that, when considering a pair \texttt{p :: (t q, t r)} of next states and the current relation \texttt{b :: [(t q, t r)]}, the pair \texttt{p} does not need to be added to the relation if it can be obtained as a combination of the elements of \texttt{b}, using the free algebra structures of \texttt{t q} and \texttt{t r}.
The first argument of \texttt{bisimT} is a function that should determine this.
Because of this abstraction, we do not actually need to constrain \texttt{t} to be a monad here.
For the \texttt{Identity} monad, one can simply use \texttt{elem} as the first argument.
The second argument is the alphabet.

Succinct automata optimized by a monad \texttt{t} enjoy a more concrete representation involving maps.
\begin{verbatim}
data SAut t a o q = SAut {
    sinitial :: t q,
    sdelta :: Map q (Map a (t q)),
    sout :: Map q o }
\end{verbatim}
This is the type of the automata that the \lstarT{} implementation learns.
The concrete representation allows the automaton to be displayed and exported.
Of course, one can determinize a succinct automaton using \texttt{t}-algebras for \texttt{a -> t q} and \texttt{o}.
\begin{verbatim}
det :: (Monad t, Ord q, Ord a) =>
    Alg t (a -> t q) -> Alg t o -> SAut t a o q -> Aut a o (t q)
\end{verbatim}
The type \texttt{Alg t x} is defined to be \texttt{t x -> x}.
We allow an arbitrary algebra on \texttt{a -> t q} rather than assuming the component \texttt{t (a -> t q) -> a -> t (t q)} of the distributive law used in earlier sections because this allows us to run the delayed monadic computations discussed earlier, which would otherwise pile up and cause serious performance issues.

\subsection{Teaching}\label{sec:imp-teach}

A teacher in our implementation is an object that comprises membership and equivalence functions.
It also records the alphabet.
\begin{verbatim}
data Teacher s a o q = Teacher {
    membership :: [a] -> s o,
    equivalence :: Aut a o q -> s (Maybe [a]),
    alphabet :: [a] }
\end{verbatim}
\texttt{Teacher} objects are parameterized by a monad \texttt{s} that serves a different purpose than optimizing the learning algorithm: it is the monad of side-effects allowed by the implementation of queries.
Whereas the \texttt{Identity} monad suffices for a predefined automaton, one may have to use the \texttt{IO} monad to interact with an actual black-box system.
By allowing an arbitrary monad rather than assuming the \texttt{IO} monad, we are able to build features such as query counters and a cache on top of any teacher through the use of \emph{monad transformers}.
A monad transformer provides for any monad a new monad into which the original one can be embedded.
For example, the \texttt{StateT x s} monad adds a state with values in \texttt{x} to an existing monad \texttt{s}.
This is the transformer that enables the addition of query counters and a cache to a teacher:
\begin{verbatim}
countTeacher :: (Monad s) =>
    Teacher s a o q -> Teacher (StateT (Int, Int) s) a o q
cacheTeacher :: (Monad s, Ord a) =>
    Teacher s a o q -> Teacher (StateT (Map [a] o) s) a o q
\end{verbatim}

The most basic teacher holds an automaton that it uses to determine membership and equivalence, the latter of which is implemented through bisimulation.
\begin{verbatim}
autTeacherT :: (Monad s, Eq o) => ((t q, t r) -> [(t q, t r)] -> Bool) ->
    [a] -> Aut a o (t q) -> Teacher s a o (t r)
\end{verbatim}
It implements a \texttt{Teacher} for any monad \texttt{s} because it does not have any side-effects.

We also provide a teacher that implements equivalence queries through random testing.
\begin{verbatim}
randomTeacher :: (Monad s, Eq o) => Int -> State StdGen [a] ->
    [a] -> ([a] -> s o) -> Teacher (StateT StdGen s) a o q
\end{verbatim}
Its first argument is the number of tests per equivalence query, while the second argument samples test words: \texttt{StdGen} is a random number generator.
Once more we use the \texttt{StateT} monad transformer, in this case to add a random number generator state to the monad \texttt{s} that the membership query function, which is the last argument, may use.
This query function is used both for membership queries and for generated test queries.
Note that this particular teacher does not give any guarantees on the validity of positive responses to equivalence queries.
We do also provide the random sampling teacher suggested by~\citet{angluin1987}, which guarantees that on a positive answer the hypothesis is \emph{probably approximately correct}, a notion introduced by~\citet{valiant1984}.
\begin{verbatim}
pacTeacher :: (Monad s, Eq o) => Double -> Double -> State StdGen [a] ->
    [a] -> ([a] -> s o) -> Teacher (StateT (Int, StdGen) s) a o q
\end{verbatim}
Here the first argument is the accuracy $\epsilon$, while the second one is the confidence $\partial$.
Both should be values between $0$ and $1$.
If $d \colon A^* \to [0, 1]$ is the distribution represented by the third argument (converting between Haskell types and sets for convenience) and $l_1, l_2 \colon A^* \to O$ are the languages of the hypothesis and the target, the guarantee is that, with probability at least $1 - \partial$, $\sum_{u \in A^*, l_1(u) \ne l_2(u)} d(u) \le \epsilon$. Compared to \texttt{randomTeacher}, an \texttt{Int} has been added to the state because the number of tests depends on the number of equivalence queries that have already been asked.

\subsection{Learning}\label{sec:imp-learn}

We define a \texttt{Learner} type that allows us to switch between variations on \lstarT{} and to optimize certain specific procedures.
\begin{verbatim}
data Learner t a o = Learner {
    decomposeRow :: ObservationTable a o -> [[a]] -> [o] -> Maybe (t [a]),
    consistencyDefect :: Maybe (ObservationTable a o -> Maybe [a]),
    ceh :: CEHandler }
\end{verbatim}
The function \texttt{decomposeRow} takes an observation table, a list of labels \texttt{l}, and a row \texttt{r}, and determines whether \texttt{r} can be obtained as a combination of the rows with labels in \texttt{l}. If this is the case, it returns the combination, which has type \texttt{t [a]}.
%
This function is used to check closedness, to minimize the labels used as states for the hypothesis, and to construct the hypothesis.
If \texttt{consistencyDefect} is set to \texttt{Nothing}, it indicates that consistency should be solved by solving closedness for what we call the \emph{transpose} of the table (swapping $S$ and $E$ and reversing their words while considering the reverse of the target language as the target language); otherwise, it contains a function that given an observation table produces a new column to fix one of its consistency defects, unless the table is already consistent.
Solving closedness for the transpose of the table always ensures consistency, but in general it may add more columns than necessary.
Lastly, \texttt{CEHandler} is a type that enumerates our adaptations of the three counterexample handling methods: the original one by~\citet{angluin1987}, the one by~\citet{maler1995}, and the one by~\citet{rivest1993}.

To enable basic implementations of \texttt{decomposeRow} and \texttt{consistencyDefect} that work for any monad $T$ (preserving finite sets), we need to be able to loop over the values of $TS$.
In order to facilitate this, there is a class \texttt{Concrete f} whose only member function turns a list of values of any type into a list of values with type \texttt{f} applied to that type.
It is intended to be the concrete application of a functor to a set (represented as a list).
We provide the functions \texttt{lazyDecomposeRow} and \texttt{lazyConsistencyDefect}, both conditioned with a \texttt{Concrete t} constraint, which directly enable a basic version of the learning algorithm.

To optimize the algorithm in a specific setting, a programmer only has to adjust these two functions.
We provide such optimized functions for the cases of non-deterministic and weighted automata (over a field).
Regarding the former case, we provide \texttt{crfsaDecompose} and \texttt{scrfsaDecompose}, which are essentially the right inverses corresponding to the canonical and simplified canonical RFSA, respectively, as explained in \examplename~\ref{ex:psefficient}.
Our optimized weighted algorithm uses Gaussian elimination in a function called \texttt{gaussianDecomposeRow} and solves consistency by solving closedness for the transpose of the table, a method readily available regardless of the monad.

Enabling our adaptation of the counterexample handling method due to Rivest and Schapire requires an additional condition.
Recall that this method requires us to pose membership queries for combinations of words, which can be done by extending the membership query function (the language) of type \texttt{[a] -> o} to one of type \texttt{t [a] -> o} using the algebra structure defined on \texttt{o}.
However, our membership query function actually has type \texttt{[a] -> s o}, and there is no reason to assume any interaction between \texttt{s} and \texttt{t}.
As a workaround, we will assume an instance of \texttt{Supported} for the monad \texttt{t}, where \texttt{Supported} is a class defined as follows:
\begin{verbatim}
class Supported f where
    supp :: (Ord a) => f a -> [a]
\end{verbatim}
Given any \texttt{u :: f a} and \texttt{g :: a -> b}, we require \texttt{supp u} to be such that the computation of \texttt{fmap g u} only evaluates \texttt{g} on the elements of \texttt{supp u}.
Naturally, we want \texttt{supp u} to be as small as possible: it should contain exactly those elements of type \texttt{a} that are present in \texttt{u}.
As an example, recall that the free semimodule monad with values in a semiring \texttt{s} can be defined on a type \texttt{a} as \texttt{Map a s}, where we identify a missing value for an element with that element being assigned zero.
Given \texttt{u :: Map a s}, \texttt{supp u} is given by the keys of the map \texttt{u} that are assigned a non-zero value. 

Using the instance for a monad \texttt{t}, the membership query function can be extended by querying the words in the support of a given element of \texttt{t [a]} sequentially, constructing a partial membership query function defined only on that support, and evaluating the extension of that function.
This method works because we assume that the side-effects exhibited by \texttt{s} do not influence future membership queries.

Finally, our general \lstarT{} implementation has the following signature:
\begin{verbatim}
lStarT :: (Monad s, Monad t, Supported t, Ord a, Eq o) =>
    Alg t (a -> t [a]) -> Alg t o ->
    Teacher s a o (t [a]) -> Learner t a o -> s (SAut t a o [a])
\end{verbatim}

\section{Experiments}\label{sec:experiments}
\definecolor{lightgray}{RGB}{220,220,220}


In this section we analyze the performance, in terms of number of queries, of several variations of our algorithm by running them on randomly generated WFAs, NFAs, and plain Moore automata. Our aim is to show the effect of exploiting the right monad and of using our adapted optimized counterexample handling method.
The experiments are run using the implementation discussed in Section~\ref{sec:implementation}.
In all cases we use an alphabet of size 3.
Random Moore automata are generated by choosing for each state an output and further for each input symbol a next state using uniform distributions.
The WFAs are over the field of size 5.
Here the outputs are chosen in the same way, and for each pair of states and each input symbol, we create a transition symbol from the first to the second state with a random weight chosen uniformly.
We take the average of 100 iterations for each of the sizes for which we generate automata.
Membership query results in tables will be rounded to whole numbers.
We use bisimulation to find counterexamples in all experiments, exploiting the fact that the target automaton will be known.
We cache membership queries so that the counts exclude duplicates.

\pgfplotstableread{experiments/dfa/lstar_mq.dat}{\dangluinmq}
\pgfplotstableread{experiments/dfa/lstarMP_mq.dat}{\dmpmq}
\pgfplotstableread{experiments/dfa/lstarRS_mq.dat}{\drsmq}
\pgfplotstableread{experiments/dfa/lstar_eq.dat}{\dangluineq}
\pgfplotstableread{experiments/dfa/lstarMP_eq.dat}{\dmpeq}
\pgfplotstableread{experiments/dfa/lstarRS_eq.dat}{\drseq}

\captionsetup[subfigure]{justification=centering}
\begin{figure}[t]%
	\centering
	\subfloat[Membership queries]{%
		\begin{tikzpicture}[every mark/.append style={mark size=1.2pt}]
			\pgfplotsset{
				ymin=0,
				xmin=20,
				xmax=200,
				width=5.5cm,
				height=4.5cm
			}
			\begin{axis}
				[
					x tick label style={/pgf/number format/1000 sep=},
					xtick={20,50,100,150,200},
					legend style={at={(1.2,1)},anchor=north west,draw=none},
					legend cell align={left},
				]
				\addplot plot table [y error index=2] {\dangluinmq};
				\addplot[mark=x,color=red] plot table [y error index=2] {\dmpmq};
				\addplot[mark=o,color=brown] plot table [y error index=2] {\drsmq};
				\legend{\lstar,\lstar$^\mathtt{MP}$,\lstar$^\mathtt{RS}$}
			\end{axis}
		\end{tikzpicture}
		\hspace{-1.8cm}
	}\hspace{1.8cm}
	\subfloat[Equivalence queries]{%
		\begin{tikzpicture}[every mark/.append style={mark size=1.2pt}]
			\pgfplotsset{
				ymin=0,
				xmin=20,
				xmax=200,
				width=5.5cm,
				height=4.5cm
			}
			\begin{axis}
				[
					x tick label style={/pgf/number format/1000 sep=},
					xtick={20,50,100,150,200},
					legend style={at={(1.03,1)},anchor=north west,draw=none},
					legend cell align={left},
				]
				\addplot plot table [y error index=2] {\dangluineq};
				\addplot[mark=x,color=red] plot table [y error index=2] {\dmpeq};
				\addplot[mark=o,color=brown] plot table [y error index=2] {\drseq};
			\end{axis}
		\end{tikzpicture}
	}
	\caption{\lstar{} variations on random DFAs.}
	\label{fig:expd}
\end{figure}

For reference, \figurename~\ref{fig:expd} compares \lstar{} and the two counterexample handling variations by Maler and Pnueli (denoted \texttt{MP}) and by Rivest and Schapire (denoted \texttt{RS}),
on randomly generated DFAs of size 20 through 200 with increments of 20.
Compared to \lstar{}, both \lstar$^\mathtt{MP}$ and \lstar$^\mathtt{RS}$ remove the need for consistency checks.
Interestingly, whereas \lstar$^\mathtt{RS}$ compared to \lstar{} improves in membership queries and worsens in equivalence queries, the situation is reversed for \lstar$^\mathtt{MP}$.

\pgfplotstableread{experiments/wfa/vangluin_mq.dat}{\wvangluinmq}
\pgfplotstableread{experiments/wfa/vmp_mq.dat}{\wvmpmq}
\pgfplotstableread{experiments/wfa/vrs_mq.dat}{\wvrsmq}
\pgfplotstableread{experiments/wfa/weighted_mq.dat}{\wweightedmq}
\pgfplotstableread{experiments/wfa/vangluin_eq.dat}{\wvangluineq}
\pgfplotstableread{experiments/wfa/vmp_eq.dat}{\wvmpeq}
\pgfplotstableread{experiments/wfa/vrs_eq.dat}{\wvrseq}
\pgfplotstableread{experiments/wfa/weighted_eq.dat}{\wweightedeq}


\subsection{\lstarT[\VMon]}

In \tablename~\ref{table:expw1} we compare the performance of \lstar{} with that of \lstarT[\VMon]{}.
(Recall that $\VMon$ is the free vector space monad.)
\begin{table}[b]
	\begin{center}
		{
			\setlength{\tabcolsep}{3pt}
			\begin{tabular}{r|r|r|r|r|r|r|r|r|r|r|r|r}
				&
					\multicolumn{6}{c|}{MQs} &
					\multicolumn{6}{c}{EQs} \\
				Size &
					\lstar &
					\lstarT[\VMon] &
					\lstar$^\mathtt{MP}$ &
					\lstarT[\VMon]$^\mathtt{MP}$ &
					\lstar$^\mathtt{RS}$ &
					\lstarT[\VMon]$^\mathtt{RS}$ &
					\lstar &
					\lstarT[\VMon] &
					\lstar$^\mathtt{MP}$ &
					\lstarT[\VMon]$^\mathtt{MP}$ &
					\lstar$^\mathtt{RS}$ &
					\lstarT[\VMon]$^\mathtt{RS}$ \\
				\hline
				1 &
					10 &
					4 &
					10 &
					4 &
					10 &
					4 &
					1.00 &
					1.00 &
					1.00 &
					1.00 &
					1.00 &
					1.00 \\
				2 &
					105 &
					15 &
					154 &
					15 &
					104 &
					11 &
					1.86 &
					1.73 &
					1.86 &
					1.73 &
					1.86 &
					1.73 \\
				3 &
					845 &
					27 &
					1003 &
					27 &
					844 &
					24 &
					2.84 &
					2.10 &
					2.16 &
					2.14 &
					3.00 &
					2.81 \\
				4 &
					5570 &
					50 &
					7904 &
					50 &
					5567 &
					40 &
					3.71 &
					2.88 &
					2.90 &
					2.83 &
					3.97 &
					3.78
			\end{tabular}
		}
	\end{center}
	\caption{\lstar{} variations and \lstarT[\VMon]{} variations on random WFAs.}
	\label{table:expw1}
\end{table}
Here \lstar{} is the obvious generalization of the original \lstar{} algorithm to learn Moore automata---DFAs with outputs in an arbitrary set, which here is the field with five elements.
Thus, as opposed to \lstarT[\VMon]{}, \lstar{} ignores the vector space structure on the output set.
In both cases we consider the three different counterexample handling methods.
The algorithms are run on randomly generated WFAs of sizes 1 through 4.
As expected, each \lstarT[\VMon]{} variation provides a massive gain over the corresponding \lstar{} variation in terms of membership queries, and a more modest one in terms of equivalence queries.
Comparing the results of the \lstar{} variations, we see that the membership query results of \lstar{} and \lstar$^\mathtt{RS}$ are extremely close together.
Other than that, the ordering of the counterexample handling methods is the same as with the DFA experiments.
The \lstarT[\VMon]{} variations will be compared in more detail later.

\pgfplotstableread{experiments/wfa/vangluinDFA_mq.dat}{\wvangluinDFAmq}
\pgfplotstableread{experiments/wfa/angluinDFA_mq.dat}{\wangluinDFAmq}
\pgfplotstableread{experiments/wfa/mpDFA_mq.dat}{\wmpDFAmq}
\pgfplotstableread{experiments/wfa/mpmDFA_mq.dat}{\wmpmDFAmq}
\pgfplotstableread{experiments/wfa/rsDFA_mq.dat}{\wrsDFAmq}
\pgfplotstableread{experiments/wfa/rsmDFA_mq.dat}{\wrsmDFAmq}
\pgfplotstableread{experiments/wfa/vangluinDFA_eq.dat}{\wvangluinDFAeq}
\pgfplotstableread{experiments/wfa/angluinDFA_eq.dat}{\wangluinDFAeq}
\pgfplotstableread{experiments/wfa/mpDFA_eq.dat}{\wmpDFAeq}
\pgfplotstableread{experiments/wfa/mpmDFA_eq.dat}{\wmpmDFAeq}
\pgfplotstableread{experiments/wfa/rsDFA_eq.dat}{\wrsDFAeq}
\pgfplotstableread{experiments/wfa/rsmDFA_eq.dat}{\wrsmDFAeq}

\captionsetup[subfigure]{justification=centering}
\begin{figure}[t]%
	\centering
	\subfloat[Membership queries]{%
		\begin{tikzpicture}[every mark/.append style={mark size=1.2pt}]
			\pgfplotsset{
				ymin=0,
				xmin=5,
				xmax=50,
				width=5.5cm,
				height=4.5cm
			}
			\begin{axis}
				[
					x tick label style={/pgf/number format/1000 sep=},
					xtick={5,20,40},
					legend style={at={(1.2,1)},anchor=north west,draw=none},
					legend cell align={left},
				]
				\addplot plot table [y error index=2] {\wangluinDFAmq};
				\addplot[mark=x,color=red] plot table [y error index=2] {\wmpDFAmq};
				\addplot[mark=o,color=brown] plot table [y error index=2] {\wrsDFAmq};
				\addplot[mark=triangle*] plot table [y error index=2] {\wvangluinDFAmq};
				\legend{\lstarT[\VMon],\lstarT[\VMon]$^\mathtt{MP}$,\lstarT[\VMon]$^\mathtt{RS}$,\lstar}
			\end{axis}
		\end{tikzpicture}
		\hspace{-1.8cm}
	}\hspace{1.8cm}
	\subfloat[Equivalence queries]{%
		\begin{tikzpicture}[every mark/.append style={mark size=1.2pt}]
			\pgfplotsset{
				ymin=0,
				xmin=5,
				xmax=50,
				width=5.5cm,
				height=4.5cm
			}
			\begin{axis}
				[
					x tick label style={/pgf/number format/1000 sep=},
					xtick={5,20,40},
					legend style={at={(1.03,1)},anchor=north west,draw=none},
					legend cell align={left},
				]
				\addplot plot table [y error index=2] {\wangluinDFAeq};
				\addplot[mark=x,color=red] plot table [y error index=2] {\wmpDFAeq};
				\addplot[mark=o,color=brown] plot table [y error index=2] {\wrsDFAeq};
				\addplot[mark=triangle*] plot table [y error index=2] {\wvangluinDFAeq};
			\end{axis}
		\end{tikzpicture}
	}
	\caption{\lstarT[\VMon]{} variations and \lstar{} on random Moore automata.}
	\label{fig:expw2}
\end{figure}

Now we run \lstar{} and variations of \lstarT[\VMon]{} on randomly generated Moore automata of sizes 5 through 50 with increments of 5.
We chose to compare the \lstarT[\VMon]{} variations only to \lstar{} because of its average performance in between \lstar$^\texttt{MP}$ and \lstar$^\texttt{RS}$ as seen in \figurename~\ref{fig:expd}.
The results are shown in \figurename~\ref{fig:expw2}.
We see that, in terms of membership queries, both \texttt{RS} and \texttt{MP} counterexample handling methods improve over the one by Angluin in this setting, and \texttt{MP} performs best in terms of either query type.
In these experiments, \lstar{} performs much better than the algorithms that attempt to take advantage of the non-existent vector space structure.
Together with the results in \tablename~\ref{table:expw1}, this is consistent with the findings of~\citet{angluin2015}: they found that for DFAs and non-deterministic, universal, and alternating automata, the adaptation of \lstar{} that takes advantage of the exact type of structure of the randomly generated target automata performs the best.

\pgfplotstableread{experiments/wfa/angluin_mq.dat}{\wangluinmq}
\pgfplotstableread{experiments/wfa/mp_mq.dat}{\wmpmq}
\pgfplotstableread{experiments/wfa/mpm_mq.dat}{\wmpmmq}
\pgfplotstableread{experiments/wfa/rs_mq.dat}{\wrsmq}
\pgfplotstableread{experiments/wfa/rsm_mq.dat}{\wrsmmq}
\pgfplotstableread{experiments/wfa/angluin_eq.dat}{\wangluineq}
\pgfplotstableread{experiments/wfa/mp_eq.dat}{\wmpeq}
\pgfplotstableread{experiments/wfa/mpm_eq.dat}{\wmpmeq}
\pgfplotstableread{experiments/wfa/rs_eq.dat}{\wrseq}
\pgfplotstableread{experiments/wfa/rsm_eq.dat}{\wrsmeq}

\captionsetup[subfigure]{justification=centering}
\begin{figure}[t]%
	\centering
	\subfloat[Membership queries]{%
		\begin{tikzpicture}[every mark/.append style={mark size=1.2pt}]
			\pgfplotsset{
				ymin=0,
				xmin=5,
				xmax=30,
				width=5.5cm,
				height=4.5cm
			}
			\begin{axis}
				[
					x tick label style={/pgf/number format/1000 sep=},
					xtick={5,10,20,30},
					legend style={at={(1.2,1)},anchor=north west,draw=none},
					legend cell align={left},
				]
				\addplot plot table [y error index=2] {\wangluinmq};
				\addplot[mark=x,color=red] plot table [y error index=2] {\wmpmq};
				\addplot[mark=o,color=brown] plot table [y error index=2] {\wrsmq};
				\legend{\lstarT[\VMon],\lstarT[\VMon]$^\mathtt{MP}$,\lstarT[\VMon]$^\mathtt{RS}$}
			\end{axis}
		\end{tikzpicture}
		\hspace{-1.8cm}
	}\hspace{1.8cm}
	\subfloat[Equivalence queries]{%
		\begin{tikzpicture}[every mark/.append style={mark size=1.2pt}]
			\pgfplotsset{
				ymin=0,
				xmin=5,
				xmax=30,
				width=5.5cm,
				height=4.5cm
			}
			\begin{axis}
				[
					x tick label style={/pgf/number format/1000 sep=},
					xtick={5,10,20,30},
					legend style={at={(1.03,1)},anchor=north west,draw=none},
					legend cell align={left},
				]
				\addplot plot table [y error index=2] {\wangluineq};
				\addplot[mark=x,color=red] plot table [y error index=2] {\wmpeq};
				\addplot[mark=o,color=brown] plot table [y error index=2] {\wrseq};
			\end{axis}
		\end{tikzpicture}
	}
	\caption{\lstarT[\VMon]{} variations on random WFAs.}
	\label{fig:expw3}
\end{figure}

\figurename~\ref{fig:expw3} illustrates the performance of \lstarT[\VMon]{} variations on randomly generated WFAs.
Here we generated WFAs of sizes 5 through 30 with increments of 5.
We emphasize that in Table~\ref{table:expw1} we could not go beyond size 4, because of performance issues with \lstar{}.
There is hardly any difference between the use of Angluin's counterexample handling method and \texttt{MP}, neither in terms of membership queries, nor in terms of equivalence queries.
Interestingly, while the \texttt{RS} method performs worse than the other methods in terms of equivalence queries, as usual, it provides no significant gain in terms of membership queries.
We ran these experiments also with the variations on the \texttt{MP} and \texttt{RS} algorithms where we drop the consistency checks.
In both cases the differences were negligible.

\pgfplotstableread{experiments/nfa/lstar_mq.dat}{\lstarmq}
\pgfplotstableread{experiments/nfa/lstarMP_mq.dat}{\lstarMPmq}
\pgfplotstableread{experiments/nfa/lstarRS_mq.dat}{\lstarRSmq}
\pgfplotstableread{experiments/nfa/nlstarMP_mq.dat}{\nlstarMPmq}
\pgfplotstableread{experiments/nfa/nlstarMPNC_mq.dat}{\nlstarMPNCmq}
\pgfplotstableread{experiments/nfa/nlstarRS_mq.dat}{\nlstarRSmq}
\pgfplotstableread{experiments/nfa/nlstarRSNC_mq.dat}{\nlstarRSNCmq}
\pgfplotstableread{experiments/nfa/lstar_eq.dat}{\lstareq}
\pgfplotstableread{experiments/nfa/lstarMP_eq.dat}{\lstarMPeq}
\pgfplotstableread{experiments/nfa/lstarRS_eq.dat}{\lstarRSeq}
\pgfplotstableread{experiments/nfa/nlstarMP_eq.dat}{\nlstarMPeq}
\pgfplotstableread{experiments/nfa/nlstarMPNC_eq.dat}{\nlstarMPNCeq}
\pgfplotstableread{experiments/nfa/nlstarRS_eq.dat}{\nlstarRSeq}
\pgfplotstableread{experiments/nfa/nlstarRSNC_eq.dat}{\nlstarRSNCeq}
\pgfplotstableread{experiments/nfa/lstar_size.dat}{\lstarsize}
\pgfplotstableread{experiments/nfa/nlstarMP_size.dat}{\nlstarMPsize}

\subsection{\nlstar}

\begin{table}[t]
	\begin{center}
			\setlength{\tabcolsep}{3pt}
			\begin{tabular}{r|r|r|r|r|r|r|r|r|r|r}
				&
					\multicolumn{5}{c|}{MQs} &
					\multicolumn{5}{c}{EQs} \\
				Size &
					\lstar &
					\nlstar$^\mathtt{MP}$ &
					\nlstar$^\mathtt{MP-}$ &
					\nlstar$^\mathtt{RS}$ &
					\nlstar$^\mathtt{RS-}$ &
					\lstar &
					\nlstar$^\mathtt{MP}$ &
					\nlstar$^\mathtt{MP-}$ &
					\nlstar$^\mathtt{RS}$ &
					\nlstar$^\mathtt{RS-}$ \\
				\hline
				4 &
					138 &
					79 &
					82 &
					55 &
					54 &
					3.75 &
					3.01 &
					3.64 &
					3.59 &
					4.64 \\
				8 &
					1792 &
					666 &
					729 &
					389 &
					381 &
					10.38 &
					6.52 &
					8.83 &
					9.37 &
					14.10 \\
				12 &
					11130 &
					2467 &
					2701 &
					1331 &
					1286 &
					18.93 &
					11.08 &
					14.92 &
					17.88 &
					27.61 \\
				16 &
					38256 &
					5699 &
					6240 &
					3036 &
					2999 &
					28.81 &
					15.75 &
					22.59 &
					27.29 &
					45.53
			\end{tabular}
	\end{center}
	\caption{\lstar{} and \nlstar{} variations on random NFAs.}
	\label{table:exp1}
\end{table}

We now consider learning algorithms for NFAs.
To generate random NFAs, we use the strategy introduced by~\citet{tabakov2005} with a transition density of $1.25$, meaning that for each input symbol there are on average $1.25$ transitions originating from each state.
According to Tabakov and Vardi, this density results in the largest equivalent minimal DFAs.
Like Tabakov and Vardi, we let half of the states be accepting.
We ran several variations of \lstar{} and \nlstar{} on randomly generated NFAs of sizes 4 through 16 with increments of 4.
The results are shown in \tablename~\ref{table:exp1}.
Here \nlstar$^\texttt{MP}$ refers to the original algorithm by~\citet{bollig2009}, with their notion of consistency; \nlstar$^\texttt{RS}$ is the same algorithm, but using the counterexample handling method that we adapted from Rivest and Schapire's.
The variations \nlstar$^\texttt{MP-}$ and \nlstar$^\texttt{RS-}$ drop the consistency checks altogether.
Unfortunately, doing the full consistency check was not computationally feasible.
As expected, the \nlstar{} algorithms yield a great improvement over \lstar{} in terms of membership queries, and in most cases they also improve in terms of equivalence queries.
This was already observed by Bollig et al.
The exception is \nlstar$^\texttt{RS-}$, which, despite having the best membership query results, requires by far the most equivalence queries.
As happened to \lstar{} on DFAs, switching within \nlstar{} from the \textsf{MP} to the \textsf{RS} counterexample handling method improves the performance in terms of membership queries and worsens it in terms of equivalence queries.
Dropping consistency altogether turns out to increase both query numbers.
\section{Conclusion}\label{sec:conclusion}

We have presented $\lstarT$, a general adaptation of \lstar{} that uses monads to learn an automaton with algebraic structure, as well as a method for finding a succinct equivalent based on its generators.
Furthermore, we adapted the optimized counterexample handling method of~\citet{rivest1993} to this setting and discussed instantiations to non-deterministic, universal, partial, weighted, alternating, and writer automata.
We have provided a prototype implementation in Haskell, using which we obtained experimental results confirming that exploiting the algebraic structure reduces the number of queries posed.
The results also reveal that the best counterexample handling method depends on the type of automata considered and the algebraic structure exploited by the algorithm.
We found that there is a significant gain in membership queries compared to the \nlstar{} algorithm by~\citet{bollig2009} when using our adapted optimized counterexample handling method.

\paragraph{Related Work.}
This paper builds on and extends the theoretical toolkit of~\citet{CALF,heerdt2016}, who are developing a categorical automata learning framework (CALF) in which learning algorithms can be understood and developed in a structured way.

An adaptation of \lstar{} that produces NFAs was first developed by~\citet{bollig2009}. Their algorithm learns a special subclass of NFAs consisting of RFSAs, which were introduced by~\citet{denis2002}. \citet{angluin2015} unified algorithms for NFAs, universal automata, and alternating automata, the latter of which was further improved by~\citet{BerndtLLR17}.
We are able to provide a more general framework, which encompasses and goes beyond those classes of automata.
Moreover, we study optimized counterexample handling, which~\citet{angluin2015,bollig2009,BerndtLLR17} do not consider.

The algorithm for weighted automata over a (not necessarily finite) field was studied in a category theoretical context by~\citet{jacobs2014} and elaborated on by~\citet{CALF}.
The algorithm itself was introduced by~\citet{bergadano1996}.
The present paper provides the first, correct-by-construction implementation of the algorithm.
The theory of succinct automata used for our hypotheses is based on the work of~\citet{arbib1975_}, revamped to more recent category theory.

Our library is currently a prototype, which is not intended to compete with a state-of-the-art tool such as LearnLib~\citep{isberner2015_} or other automata learning libraries like libalf~\citep{bollig2010}.
Our Haskell implementation does not provide the computational efficiency achieved by LearnLib, which furthermore includes the TTT-algorithm with its optimized data structure that 
replaces the observation table by a tree~\citep{isberner2014}.
Such optimization is ad-hoc for DFAs, and an extension to other classes of automata is not trivial. First steps in this direction have been done by~\citep{CALF}, who have studied the tree data structure in a more general setting.
We intend to further pursue investigation in this direction, in order to allow for optimized data structures in a future version of our library.
We note that, although libalf supports NFAs, none of the existing tools and libraries offers the flexibility of our library, in terms of available optimizations and classes of models that can be learned.


%
\paragraph{Future Work.}
Whereas our general algorithm effortlessly instantiates to monads that preserve finite sets, a major challenge lies in investigating monads that do not enjoy this property. In fact, although the algorithm for weighted automata generalizes to an infinite field~\citep{jacobs2014,CALF}, for an infinite semiring in general we cannot guarantee termination.
This is because a finitely generated semimodule may have an infinite chain of strict submodules.
Intuitively, this means that while fixing closedness defects increases the size of the hypothesis state space semimodule, an infinite number of steps may be needed to resolve all closedness defects.
There are however subclasses of semirings for which a generalization should be possible, e.g., Noetherian or, more generally, proper semirings, which were recently studied by~\citet{Milius2017}.
%
%
 Moreover, we expect that \lstarT{} can be generalized from the category of sets to \emph{locally finitely presentable} categories. 

As a result of the correspondence between learning and conformance testing~\citep{berg2005,CALF}, it should be possible to include in our framework the W-method~\citep{chow1978}, which is often used in case studies deploying \lstar{}~\citep[e.g.,][]{chalupar2014,ruiter2015}.
We defer a thorough investigation of conformance testing to future work.


%

\bibliography{main}


\begin{thebibliography}{37}


\ifx \showCODEN    \undefined \def \showCODEN     #1{\unskip}     \fi
\ifx \showDOI      \undefined \def \showDOI       #1{#1}\fi
\ifx \showISBNx    \undefined \def \showISBNx     #1{\unskip}     \fi
\ifx \showISBNxiii \undefined \def \showISBNxiii  #1{\unskip}     \fi
\ifx \showISSN     \undefined \def \showISSN      #1{\unskip}     \fi
\ifx \showLCCN     \undefined \def \showLCCN      #1{\unskip}     \fi
\ifx \shownote     \undefined \def \shownote      #1{#1}          \fi
\ifx \showarticletitle \undefined \def \showarticletitle #1{#1}   \fi
\ifx \showURL      \undefined \def \showURL       {\relax}        \fi
\providecommand\bibfield[2]{#2}
\providecommand\bibinfo[2]{#2}
\providecommand\natexlab[1]{#1}
\providecommand\showeprint[2][]{arXiv:#2}

\bibitem[\protect\citeauthoryear{Aarts, de~Ruiter, and Poll}{Aarts
  et~al\mbox{.}}{2013}]%
        {aarts2013}
\bibfield{author}{\bibinfo{person}{Fides Aarts}, \bibinfo{person}{Joeri de
  Ruiter}, {and} \bibinfo{person}{Erik Poll}.} \bibinfo{year}{2013}\natexlab{}.
\newblock \showarticletitle{Formal models of bank cards for free}. In
  \bibinfo{booktitle}{\emph{ICSTW}}. IEEE Computer Society,
  \bibinfo{pages}{461--468}.
\newblock


\bibitem[\protect\citeauthoryear{Aminof, Kupferman, and Lampert}{Aminof
  et~al\mbox{.}}{2010}]%
        {AminofKL10}
\bibfield{author}{\bibinfo{person}{Benjamin Aminof}, \bibinfo{person}{Orna
  Kupferman}, {and} \bibinfo{person}{Robby Lampert}.}
  \bibinfo{year}{2010}\natexlab{}.
\newblock \showarticletitle{Reasoning about online algorithms with weighted
  automata}.
\newblock \bibinfo{journal}{\emph{{ACM} Trans. Algorithms}}
  \bibinfo{volume}{6}, \bibinfo{number}{2} (\bibinfo{year}{2010}),
  \bibinfo{pages}{28:1--28:36}.
\newblock


\bibitem[\protect\citeauthoryear{Angluin}{Angluin}{1987}]%
        {angluin1987}
\bibfield{author}{\bibinfo{person}{Dana Angluin}.}
  \bibinfo{year}{1987}\natexlab{}.
\newblock \showarticletitle{Learning Regular Sets from Queries and
  Counterexamples}.
\newblock \bibinfo{journal}{\emph{Inform. Comput.}}  \bibinfo{volume}{75}
  (\bibinfo{year}{1987}), \bibinfo{pages}{87--106}.
\newblock


\bibitem[\protect\citeauthoryear{Angluin, Eisenstat, and Fisman}{Angluin
  et~al\mbox{.}}{2015}]%
        {angluin2015}
\bibfield{author}{\bibinfo{person}{Dana Angluin}, \bibinfo{person}{Sarah
  Eisenstat}, {and} \bibinfo{person}{Dana Fisman}.}
  \bibinfo{year}{2015}\natexlab{}.
\newblock \showarticletitle{Learning regular languages via alternating
  automata}. In \bibinfo{booktitle}{\emph{IJCAI}}. \bibinfo{pages}{3308--3314}.
\newblock


\bibitem[\protect\citeauthoryear{Arbib and Manes}{Arbib and Manes}{1975}]%
        {arbib1975_}
\bibfield{author}{\bibinfo{person}{Michael~A. Arbib} {and}
  \bibinfo{person}{Ernest~G. Manes}.} \bibinfo{year}{1975}\natexlab{}.
\newblock \showarticletitle{Fuzzy machines in a category}.
\newblock \bibinfo{journal}{\emph{Bulletin of the AMS}}  \bibinfo{volume}{13}
  (\bibinfo{year}{1975}), \bibinfo{pages}{169--210}.
\newblock


\bibitem[\protect\citeauthoryear{Baier, Gr{\"o}{\ss}er, and Ciesinski}{Baier
  et~al\mbox{.}}{2009}]%
        {BaierGC09}
\bibfield{author}{\bibinfo{person}{Christel Baier}, \bibinfo{person}{Marcus
  Gr{\"o}{\ss}er}, {and} \bibinfo{person}{Frank Ciesinski}.}
  \bibinfo{year}{2009}\natexlab{}.
\newblock \showarticletitle{Model checking linear-time properties of
  probabilistic systems}. In \bibinfo{booktitle}{\emph{Handbook of Weighted
  automata}}.
\newblock


\bibitem[\protect\citeauthoryear{Berg, Grinchtein, Jonsson, Leucker, Raffelt,
  and Steffen}{Berg et~al\mbox{.}}{2005}]%
        {berg2005}
\bibfield{author}{\bibinfo{person}{Therese Berg}, \bibinfo{person}{Olga
  Grinchtein}, \bibinfo{person}{Bengt Jonsson}, \bibinfo{person}{Martin
  Leucker}, \bibinfo{person}{Harald Raffelt}, {and} \bibinfo{person}{Bernhard
  Steffen}.} \bibinfo{year}{2005}\natexlab{}.
\newblock \showarticletitle{On the correspondence between conformance testing
  and regular inference}. In \bibinfo{booktitle}{\emph{FASE}},
  Vol.~\bibinfo{volume}{3442}. \bibinfo{pages}{175--189}.
\newblock


\bibitem[\protect\citeauthoryear{Bergadano and Varricchio}{Bergadano and
  Varricchio}{1996}]%
        {bergadano1996}
\bibfield{author}{\bibinfo{person}{Francesco Bergadano} {and}
  \bibinfo{person}{Stefano Varricchio}.} \bibinfo{year}{1996}\natexlab{}.
\newblock \showarticletitle{Learning behaviors of automata from multiplicity
  and equivalence queries}.
\newblock \bibinfo{journal}{\emph{SIAM J. Comput.}}  \bibinfo{volume}{25}
  (\bibinfo{year}{1996}), \bibinfo{pages}{1268--1280}.
\newblock


\bibitem[\protect\citeauthoryear{Berndt, Li{\'{s}}kiewicz, Lutter, and
  Reischuk}{Berndt et~al\mbox{.}}{2017}]%
        {BerndtLLR17}
\bibfield{author}{\bibinfo{person}{Sebastian Berndt}, \bibinfo{person}{Maciej
  Li{\'{s}}kiewicz}, \bibinfo{person}{Matthias Lutter}, {and}
  \bibinfo{person}{R{\"{u}}diger Reischuk}.} \bibinfo{year}{2017}\natexlab{}.
\newblock \showarticletitle{Learning Residual Alternating Automata}. In
  \bibinfo{booktitle}{\emph{AAAI}}. \bibinfo{pages}{1749--1755}.
\newblock


\bibitem[\protect\citeauthoryear{Bertrand}{Bertrand}{2017}]%
        {Meven}
\bibfield{author}{\bibinfo{person}{Meven Bertrand}.}
  \bibinfo{year}{2017}\natexlab{}.
\newblock \showarticletitle{Coalgebraic Determinization of Alternating
  Automata}.
\newblock  (\bibinfo{year}{2017}).
\newblock
\newblock
\shownote{\url{http://jurriaan.creativecode.org/wp-content/uploads/2017/10/alt.pdf}.}


\bibitem[\protect\citeauthoryear{Bollig, Habermehl, Kern, and Leucker}{Bollig
  et~al\mbox{.}}{2008}]%
        {Bollig09TR}
\bibfield{author}{\bibinfo{person}{Benedikt Bollig}, \bibinfo{person}{Peter
  Habermehl}, \bibinfo{person}{Carsten Kern}, {and} \bibinfo{person}{Martin
  Leucker}.} \bibinfo{year}{2008}\natexlab{}.
\newblock \bibinfo{booktitle}{\emph{Angluin-Style Learning of NFA (Research
  Report LSV-08-28)}}.
\newblock \bibinfo{type}{{T}echnical {R}eport}. \bibinfo{institution}{ENS
  Cachan}.
\newblock


\bibitem[\protect\citeauthoryear{Bollig, Habermehl, Kern, and Leucker}{Bollig
  et~al\mbox{.}}{2009}]%
        {bollig2009}
\bibfield{author}{\bibinfo{person}{Benedikt Bollig}, \bibinfo{person}{Peter
  Habermehl}, \bibinfo{person}{Carsten Kern}, {and} \bibinfo{person}{Martin
  Leucker}.} \bibinfo{year}{2009}\natexlab{}.
\newblock \showarticletitle{Angluin-Style Learning of {NFA}}. In
  \bibinfo{booktitle}{\emph{IJCAI}}, Vol.~\bibinfo{volume}{9}.
  \bibinfo{pages}{1004--1009}.
\newblock


\bibitem[\protect\citeauthoryear{Bollig, Katoen, Kern, Leucker, Neider, and
  Piegdon}{Bollig et~al\mbox{.}}{2010}]%
        {bollig2010}
\bibfield{author}{\bibinfo{person}{Benedikt Bollig},
  \bibinfo{person}{Joost-Pieter Katoen}, \bibinfo{person}{Carsten Kern},
  \bibinfo{person}{Martin Leucker}, \bibinfo{person}{Daniel Neider}, {and}
  \bibinfo{person}{David~R Piegdon}.} \bibinfo{year}{2010}\natexlab{}.
\newblock \showarticletitle{libalf: The automata learning framework}. In
  \bibinfo{booktitle}{\emph{CAV}}. \bibinfo{pages}{360--364}.
\newblock


\bibitem[\protect\citeauthoryear{Chalupar, Peherstorfer, Poll, and
  de~Ruiter}{Chalupar et~al\mbox{.}}{2014}]%
        {chalupar2014}
\bibfield{author}{\bibinfo{person}{Georg Chalupar}, \bibinfo{person}{Stefan
  Peherstorfer}, \bibinfo{person}{Erik Poll}, {and} \bibinfo{person}{Joeri de
  Ruiter}.} \bibinfo{year}{2014}\natexlab{}.
\newblock \showarticletitle{Automated Reverse Engineering using
  {L}ego{\textregistered}}. In \bibinfo{booktitle}{\emph{WOOT}}.
\newblock


\bibitem[\protect\citeauthoryear{Chatterjee, Doyen, and Henzinger}{Chatterjee
  et~al\mbox{.}}{2008}]%
        {ChatterjeeDH08}
\bibfield{author}{\bibinfo{person}{Krishnendu Chatterjee},
  \bibinfo{person}{Laurent Doyen}, {and} \bibinfo{person}{Thomas~A.
  Henzinger}.} \bibinfo{year}{2008}\natexlab{}.
\newblock \showarticletitle{Quantitative Languages}. In
  \bibinfo{booktitle}{\emph{{CSL}}}. \bibinfo{pages}{385--400}.
\newblock


\bibitem[\protect\citeauthoryear{Cho, Babi{\'c}, Shin, and Song}{Cho
  et~al\mbox{.}}{2010}]%
        {cho2010}
\bibfield{author}{\bibinfo{person}{Chia~Yuan Cho}, \bibinfo{person}{Domagoj
  Babi{\'c}}, \bibinfo{person}{Eui Chul~Richard Shin}, {and}
  \bibinfo{person}{Dawn Song}.} \bibinfo{year}{2010}\natexlab{}.
\newblock \showarticletitle{Inference and Analysis of Formal Models of Botnet
  Command and Control Protocols}. In \bibinfo{booktitle}{\emph{CCS}}. ACM,
  \bibinfo{pages}{426--439}.
\newblock


\bibitem[\protect\citeauthoryear{Chow}{Chow}{1978}]%
        {chow1978}
\bibfield{author}{\bibinfo{person}{Tsun~S. Chow}.}
  \bibinfo{year}{1978}\natexlab{}.
\newblock \showarticletitle{Testing Software Design Modeled by Finite-State
  Machines}.
\newblock \bibinfo{journal}{\emph{IEEE Trans. Software Eng.}}
  \bibinfo{volume}{4} (\bibinfo{year}{1978}), \bibinfo{pages}{178--187}.
\newblock


\bibitem[\protect\citeauthoryear{de~Ruiter and Poll}{de~Ruiter and
  Poll}{2015}]%
        {ruiter2015}
\bibfield{author}{\bibinfo{person}{Joeri de Ruiter} {and} \bibinfo{person}{Erik
  Poll}.} \bibinfo{year}{2015}\natexlab{}.
\newblock \showarticletitle{Protocol state fuzzing of {TLS} implementations}.
  In \bibinfo{booktitle}{\emph{USENIX Security}}. \bibinfo{pages}{193--206}.
\newblock


\bibitem[\protect\citeauthoryear{Denis, Lemay, and Terlutte}{Denis
  et~al\mbox{.}}{2002}]%
        {denis2002}
\bibfield{author}{\bibinfo{person}{Fran{\c{c}}ois Denis},
  \bibinfo{person}{Aur{\'e}lien Lemay}, {and} \bibinfo{person}{Alain
  Terlutte}.} \bibinfo{year}{2002}\natexlab{}.
\newblock \showarticletitle{Residual finite state automata}.
\newblock \bibinfo{journal}{\emph{Fundamenta Informaticae}}
  \bibinfo{volume}{51} (\bibinfo{year}{2002}), \bibinfo{pages}{339--368}.
\newblock


\bibitem[\protect\citeauthoryear{Droste and Gastin}{Droste and Gastin}{2005}]%
        {DrosteG05}
\bibfield{author}{\bibinfo{person}{Manfred Droste} {and} \bibinfo{person}{Paul
  Gastin}.} \bibinfo{year}{2005}\natexlab{}.
\newblock \showarticletitle{Weighted Automata and Weighted Logics}. In
  \bibinfo{booktitle}{\emph{{ICALP}}}. \bibinfo{pages}{513--525}.
\newblock


\bibitem[\protect\citeauthoryear{Isberner, Howar, and Steffen}{Isberner
  et~al\mbox{.}}{2014}]%
        {isberner2014}
\bibfield{author}{\bibinfo{person}{Malte Isberner}, \bibinfo{person}{Falk
  Howar}, {and} \bibinfo{person}{Bernhard Steffen}.}
  \bibinfo{year}{2014}\natexlab{}.
\newblock \showarticletitle{The {TTT} algorithm: A redundancy-free approach to
  active automata learning}. In \bibinfo{booktitle}{\emph{Runtime
  Verification}} \emph{(\bibinfo{series}{LNCS})}, Vol.~\bibinfo{volume}{8734}.
  \bibinfo{pages}{307--322}.
\newblock


\bibitem[\protect\citeauthoryear{Isberner, Howar, and Steffen}{Isberner
  et~al\mbox{.}}{2015}]%
        {isberner2015_}
\bibfield{author}{\bibinfo{person}{Malte Isberner}, \bibinfo{person}{Falk
  Howar}, {and} \bibinfo{person}{Bernhard Steffen}.}
  \bibinfo{year}{2015}\natexlab{}.
\newblock \showarticletitle{The Open-Source {L}earn{L}ib}. In
  \bibinfo{booktitle}{\emph{CAV}} \emph{(\bibinfo{series}{LNCS})},
  Vol.~\bibinfo{volume}{9206}. \bibinfo{pages}{487--495}.
\newblock


\bibitem[\protect\citeauthoryear{Jacobs and Silva}{Jacobs and Silva}{2014}]%
        {jacobs2014}
\bibfield{author}{\bibinfo{person}{Bart Jacobs} {and}
  \bibinfo{person}{Alexandra Silva}.} \bibinfo{year}{2014}\natexlab{}.
\newblock \showarticletitle{Automata Learning: A Categorical Perspective}. In
  \bibinfo{booktitle}{\emph{Horizons of the Mind}},
  Vol.~\bibinfo{volume}{8464}. \bibinfo{pages}{384--406}.
\newblock


\bibitem[\protect\citeauthoryear{Kiselyov and Ishii}{Kiselyov and
  Ishii}{2015}]%
        {kiselyov2015}
\bibfield{author}{\bibinfo{person}{Oleg Kiselyov} {and} \bibinfo{person}{Hiromi
  Ishii}.} \bibinfo{year}{2015}\natexlab{}.
\newblock \showarticletitle{Freer monads, more extensible effects}. In
  \bibinfo{booktitle}{\emph{ACM SIGPLAN Notices}}, Vol.~\bibinfo{volume}{50}.
  ACM, \bibinfo{pages}{94--105}.
\newblock


\bibitem[\protect\citeauthoryear{Kozen}{Kozen}{2012}]%
        {kozen-book}
\bibfield{author}{\bibinfo{person}{Dexter~C. Kozen}.}
  \bibinfo{year}{2012}\natexlab{}.
\newblock \bibinfo{booktitle}{\emph{Automata and computability}}.
\newblock \bibinfo{publisher}{Springer Science \& Business Media}.
\newblock


\bibitem[\protect\citeauthoryear{Kuperberg}{Kuperberg}{2014}]%
        {Kuperberg14}
\bibfield{author}{\bibinfo{person}{Denis Kuperberg}.}
  \bibinfo{year}{2014}\natexlab{}.
\newblock \showarticletitle{Linear Temporal Logic for Regular Cost Functions}.
\newblock \bibinfo{journal}{\emph{Logical Methods in Computer Science}}
  \bibinfo{volume}{10}, \bibinfo{number}{1} (\bibinfo{year}{2014}).
\newblock


\bibitem[\protect\citeauthoryear{Maler and Pnueli}{Maler and Pnueli}{1995}]%
        {maler1995}
\bibfield{author}{\bibinfo{person}{Oded Maler} {and} \bibinfo{person}{Amir
  Pnueli}.} \bibinfo{year}{1995}\natexlab{}.
\newblock \showarticletitle{On the Learnability of Infinitary Regular Sets}.
\newblock \bibinfo{journal}{\emph{Inform. and Comput.}}  \bibinfo{volume}{118}
  (\bibinfo{year}{1995}), \bibinfo{pages}{316--326}.
\newblock


\bibitem[\protect\citeauthoryear{Milius}{Milius}{2017}]%
        {Milius2017}
\bibfield{author}{\bibinfo{person}{Stefan Milius}.}
  \bibinfo{year}{2017}\natexlab{}.
\newblock \showarticletitle{Proper functors}.
\newblock  (\bibinfo{year}{2017}).
\newblock
\newblock
\shownote{Submitted, copy obtained in personal communication.}


\bibitem[\protect\citeauthoryear{Rivest and Schapire}{Rivest and
  Schapire}{1993}]%
        {rivest1993}
\bibfield{author}{\bibinfo{person}{Ronald~L. Rivest} {and}
  \bibinfo{person}{Robert~E. Schapire}.} \bibinfo{year}{1993}\natexlab{}.
\newblock \showarticletitle{Inference of Finite Automata Using Homing
  Sequences}.
\newblock \bibinfo{journal}{\emph{Inform. Comput.}}  \bibinfo{volume}{103}
  (\bibinfo{year}{1993}), \bibinfo{pages}{299--347}.
\newblock


\bibitem[\protect\citeauthoryear{Rot, Bonsangue, and Rutten}{Rot
  et~al\mbox{.}}{2013}]%
        {rot2013}
\bibfield{author}{\bibinfo{person}{Jurriaan Rot}, \bibinfo{person}{Marcello
  Bonsangue}, {and} \bibinfo{person}{Jan Rutten}.}
  \bibinfo{year}{2013}\natexlab{}.
\newblock \showarticletitle{Coalgebraic bisimulation-up-to}. In
  \bibinfo{booktitle}{\emph{SOFSEM}}. \bibinfo{pages}{369--381}.
\newblock


\bibitem[\protect\citeauthoryear{Sangiorgi}{Sangiorgi}{1998}]%
        {sangiorgi1998}
\bibfield{author}{\bibinfo{person}{Davide Sangiorgi}.}
  \bibinfo{year}{1998}\natexlab{}.
\newblock \showarticletitle{On the bisimulation proof method}.
\newblock \bibinfo{journal}{\emph{Mathematical Structures in Computer Science}}
   \bibinfo{volume}{8} (\bibinfo{year}{1998}), \bibinfo{pages}{447--479}.
\newblock


\bibitem[\protect\citeauthoryear{Schuts, Hooman, and Vaandrager}{Schuts
  et~al\mbox{.}}{2016}]%
        {schuts2016}
\bibfield{author}{\bibinfo{person}{Mathijs Schuts}, \bibinfo{person}{Jozef
  Hooman}, {and} \bibinfo{person}{Frits Vaandrager}.}
  \bibinfo{year}{2016}\natexlab{}.
\newblock \showarticletitle{Refactoring of legacy software using model learning
  and equivalence checking: an industrial experience report}. In
  \bibinfo{booktitle}{\emph{IFM}}, Vol.~\bibinfo{volume}{9681}.
  \bibinfo{pages}{311--325}.
\newblock


\bibitem[\protect\citeauthoryear{Tabakov and Vardi}{Tabakov and Vardi}{2005}]%
        {tabakov2005}
\bibfield{author}{\bibinfo{person}{Deian Tabakov} {and}
  \bibinfo{person}{Moshe~Y Vardi}.} \bibinfo{year}{2005}\natexlab{}.
\newblock \showarticletitle{Experimental evaluation of classical automata
  constructions}. In \bibinfo{booktitle}{\emph{LPAR}},
  Vol.~\bibinfo{volume}{5}. \bibinfo{pages}{396--411}.
\newblock


\bibitem[\protect\citeauthoryear{Vaandrager}{Vaandrager}{2017}]%
        {cacm}
\bibfield{author}{\bibinfo{person}{Frits~W. Vaandrager}.}
  \bibinfo{year}{2017}\natexlab{}.
\newblock \showarticletitle{Model learning}.
\newblock \bibinfo{journal}{\emph{Commun. {ACM}}} \bibinfo{volume}{60},
  \bibinfo{number}{2} (\bibinfo{year}{2017}), \bibinfo{pages}{86--95}.
\newblock


\bibitem[\protect\citeauthoryear{Valiant}{Valiant}{1984}]%
        {valiant1984}
\bibfield{author}{\bibinfo{person}{Leslie~G. Valiant}.}
  \bibinfo{year}{1984}\natexlab{}.
\newblock \showarticletitle{A theory of the learnable}.
\newblock \bibinfo{journal}{\emph{Commun. ACM}}  \bibinfo{volume}{27}
  (\bibinfo{year}{1984}), \bibinfo{pages}{1134--1142}.
\newblock


\bibitem[\protect\citeauthoryear{van Heerdt}{van Heerdt}{2016}]%
        {heerdt2016}
\bibfield{author}{\bibinfo{person}{Gerco van Heerdt}.}
  \bibinfo{year}{2016}\natexlab{}.
\newblock \emph{\bibinfo{title}{An Abstract Automata Learning Framework}}.
\newblock \bibinfo{thesistype}{Master's\ thesis}. \bibinfo{school}{Radboud
  University Nijmegen}.
\newblock


\bibitem[\protect\citeauthoryear{van Heerdt, Sammartino, and Silva}{van Heerdt
  et~al\mbox{.}}{2017}]%
        {CALF}
\bibfield{author}{\bibinfo{person}{Gerco van Heerdt}, \bibinfo{person}{Matteo
  Sammartino}, {and} \bibinfo{person}{Alexandra Silva}.}
  \bibinfo{year}{2017}\natexlab{}.
\newblock \showarticletitle{{CALF:} Categorical Automata Learning Framework}.
  In \bibinfo{booktitle}{\emph{{CSL}}}. \bibinfo{pages}{29:1--29:24}.
\newblock


\end{thebibliography}

\clearpage
\appendix

\section{Omitted proofs}

\finitet*
\begin{proof}
The left to right implication is proved by freely generating a $T$-automaton from the Moore one via the monad unit, and by recalling that $T$ preserves finite sets. The resulting $T$-automaton accepts $\lang$ and is finite, therefore any of its quotients, including the minimal $T$-automaton accepting $\lang$, is finite. Analogously, the right to left implication follows by forgetting the algebraic structure of the $T$-automaton: this yields a finite Moore automaton accepting $\lang$.
\end{proof}

\succlang*
\begin{proof}
	Assume a right inverse $i \colon H \to T(S')$ of $e \colon T(S') \to H$.
	We first prove $o_\hyp \circ e = o_\suc$, by induction on the length of words.
	For all $U \in T(S')$, we have
	\begin{align*}
		o_\hyp(e(U))(\eword) &
			= \out_\hyp(e(U)) &
			&
			\text{(definition of $o_\hyp$)} \\
		&
			= \out_\hyp(\Trow^\sharp(U)) &
			&
			\text{(definition of $e$)} \\
		&
			= \Trow^\sharp(U)(\eword) &
			&
			\text{(definition of $\out_\hyp$)} \\
		&
			= \out_\suc(U) &
			&
			\text{(definition of $\out_\suc$)} \\
		&
			= o_\suc(U)(\eword) &
			&
			\text{(definition of $o_\suc$)}.
	\end{align*}
	Now assume that for a given $v \in A^*$ and all $U \in T(S')$ we have $o_\hyp(e(U))(v) = o_\suc(U)(v)$.
	Then, for all $U \in T(S')$ and $a \in A$,
	\begin{align*}
		o_\hyp(e(U))(av) &
			= o_\hyp(\delta_\hyp(e(U))(a))(v) &
			&
			\text{(definition of $o_\hyp$)} \\
		&
			= o_\hyp(\delta_\hyp(\Trow^\sharp(U))(a))(v) &
			&
			\text{(definition of $e$)} \\
		&
			= o_\hyp(\Brow^\sharp(U)(a))(v) &
			&
			\text{(definition of $\delta_\hyp$)} \\
		&
			= (o_\hyp \circ e \circ i)(\Brow^\sharp(U)(a))(v) &
			&
			\text{($e \circ i = \idf_H$)} \\
		&
			= (o_\suc \circ i)(\Brow^\sharp(U)(a))(v) &
			&
			\text{(induction hypothesis)} \\
		&
			= o_\suc(\delta_\suc(U)(a))(v) &
			&
			\text{(definition of $\delta_\suc$)} \\
		&
			= o_\suc(U)(av) &
			&
			\text{(definition of $o_\suc$)}.
	\end{align*}
	From this we see that
	\begin{align*}
		o_\suc(\init_\suc) &
			= (o_\suc \circ i \circ \Trow)(\eword) &
			&
			\text{(definition of $\init_\suc$)} \\
		&
			= (o_\hyp \circ e \circ i \circ \Trow)(\eword) &
			&
			\text{($o_\hyp \circ e = o_\suc$)} \\
		&
			= (o_\hyp \circ \Trow)(\eword) &
			&
			\text{($e \circ i = \idf_H$)} \\
		&
			= o_\hyp(\init_\hyp) &
			&
			\text{(definition of $\init_\hyp$)}.
		\qedhere
	\end{align*}
\end{proof}

\genalg*
\begin{proof}
	Minimality is obvious, as $S'$ not being minimal would make the loop guard true.

	We prove that the returned set is a set of generators.
	For clarity, we denote by $d_{S'} \colon S \to T(S')$ the function associated with a set of generators $S'$.
	The main idea is incrementally building $d_{S'}$ while building $S'$.
	In the first line, $S$ is a set of generators, with $d_S = \eta_S \colon S \to T(S)$.
	For the loop, suppose $S'$ is a set of generators.
	If the loop guard is false, the algorithm returns the set of generators $S'$.
	Otherwise, suppose there are there are $s \in S'$ and $U \in T(S' \setminus \{s\})$ such that $\Trow^\sharp(U) = \Trow(s)$.
	Then there is a function
	\begin{align*}
		f \colon S' \to T(S' \setminus \{s\}) &
			&
			f(s') = \begin{cases}
				U &
					\text{if $s' = s$} \\
				\eta(s') &
					\text{if $s' \ne s$}
			\end{cases}
	\end{align*}
	that satisfies $\Trow(s') = \Trow^\sharp(f(s'))$ for all $s' \in S'$, from which it follows that $\Trow^\sharp(U') = \Trow^\sharp(f^\sharp(U'))$ for all $U' \in T(S')$.
	Then we can set $d_{S' \setminus \{s\}}$ to $f^\sharp \circ d_{S'} \colon S \to T(S' \setminus \{s\})$ because $\Trow(s') = \Trow^\sharp(d_{S' \setminus \{s\}}(s'))$ for all $s' \in S$.
	Therefore, $S' \setminus \{s\}$ is a set of generators.
\end{proof}

\end{document}